\documentclass[twocolumn]{aastex63}

\usepackage{xspace}
\usepackage{appendix}
\usepackage[version=4]{mhchem}
\ExplSyntaxOn
\keys_define:nn { mhchem}
 {
  arrow-min-length .code:n =
   \cs_set:Npn \__mhchem_arrow_options_minLength:n { {#1}} 
}
\ExplSyntaxOff

\newcommand{\s}{\ensuremath{{\,\rm s^{-1}}}\xspace}

\newcommand{\chtnc}{CH$_{3}$NC\xspace}
\newcommand{\chtcn}{CH$_{3}$CN\xspace}
\newcommand{\ctn}{C$_3$N\xspace}
\newcommand{\hctn}{HC$_3$N\xspace}
\newcommand{\av}{$A_{\rm{V}}$\xspace}

\newcommand{\ccc}{\ensuremath{{\,\rm cm^{-3}}}\xspace}

\received{July 30th, 2019}
\revised{October 1st, 2019}
\accepted{October 2nd, 2019}

\shorttitle{The role of C/O in nitrile astrochemistry in PDRs and planet-forming disks}
\shortauthors{Le Gal et al.}

\begin{document}

\title{The role of C/O in nitrile astrochemistry in PDRs and planet-forming disks}


\correspondingauthor{Romane Le Gal}
\email{romane.le$\_$gal@cfa.harvard.edu}

\author[0000-0003-1837-3772]{Romane Le Gal}
\affiliation{Harvard-Smithsonian Center for Astrophysics, 60 Garden St., Cambridge, MA 02138, USA}
\author{Madison~T.~Brady}
\altaffiliation{Noland Internship at Harvard-Smithsonian Center for Astrophysics (summer 2018)}
\affiliation{California Institute of Technology, Pasadena, CA 91125, USA}

\author[0000-0001-8798-1347]{Karin I. \"Oberg}
\affiliation{Harvard-Smithsonian Center for Astrophysics, 60 Garden St., Cambridge, MA 02138, USA}
\author[0000-0002-4949-8562]{Evelyne~Roueff}
\author{Franck~Le Petit}
\affiliation{Sorbonne Universit\'e, Observatoire de Paris, Universit\'e PSL, CNRS, LERMA, F-92190,  Meudon, France}

\begin{abstract}
Complex nitriles, such as \hctn, and \chtcn, are observed in a wide variety of astrophysical environments, including at relatively high abundances in photon-dominated regions (PDR) and the UV exposed atmospheres of planet-forming disks. The latter have been inferred to be oxygen-poor, suggesting that these observations may be explained by organic chemistry in C-rich environments. In this study we first explore if the PDR complex nitrile observations can be explained by gas-phase PDR chemistry alone if the elemental C/O ratio is elevated.   
In the case of the Horsehead PDR, we find that gas-phase chemistry with C/O $\gtrsim$ 0.9 can indeed explain the observed nitrile abundances, increasing predicted abundances by several orders of magnitude compared to standard C/O assumptions. We also find that the nitrile abundances are sensitive to the cosmic ray ionization treatment, and provide constraints on the branching ratios between \chtcn and \ce{CH3NC} productions. In a fiducial disk model, an elevated C/O ratio increases the \chtcn and \hctn productions by more than an order of magnitude, bringing abundance predictions within an order of magnitude to what has been inferred from observations. The C/O ratio appears to be a key variable in predicting and interpreting complex organic molecule abundances in photon-dominated regions across a range of scales. 
\end{abstract}


\section{Introduction}
\label{sec:intro}

Origins of life on Earth must have been closely linked to the emergence of information-rich polymers such as DNA (deoxyribonucleic acid) or RNA (ribonucleic acid). While their initial formation on the early Earth remains mysterious, there are plausible chemical pathways to their building blocks on the early Earth through a nitrile-centered UV-driven chemistry \citep{Powner2009,patel2015,sutherland2016}. Simple and complex nitriles are abundantly found at all stages of star and planet formation, including in planet-forming disks, suggesting that the organic chemistry that preceded life on Earth is not unique to the Solar System \citep{chapillon2012,oberg2015nat,bergner2018,loomis2018}. 
Perhaps surprisingly \chtcn and HC$_3$N are two of the most commonly detected larger organic molecules in disks, and the origins of these high abundances are uncertain. \cite{oberg2015nat} and \citet{loomis2018} both invoke grain-surface chemical pathways to predict sufficient amounts of \chtcn, but these predictions are extremely uncertain due to lack of experimental data on ice nitrile chemistry and desorption.

One important observational constraint is that observed \hctn and \chtcn emissions appear to come from the upper most layer of disks or disk atmospheres \citep{oberg2015nat,bergner2018,loomis2018}. Disk atmospheres are proposed analogs to the more well-studied photon-dominated regions (PDR). Interestingly, complex nitriles have also been detected at unexpectedly high abundances in the deeply characterized PDR, the Horsehead nebula \citep{gratier2013}. Located in the Orion constellation and seen almost edge-on \citep{abergel2003}, the Horsehead nebula constitutes a perfect template source to study in detail the physics and chemistry occurring in PDRs. With the WHISPER survey\footnote{http://www.iram-institute.org/$\sim$horsehead/\\Horsehead\_Nebula/WHISPER.html} (Wideband High-resolution Iram-30m Survey at two Positions with Emir Receivers, PI: J. Pety), the chemistry of this PDR has been surveyed at unprecedented detail, both at the edge of the PDR \citep[defined by the HCO peak emission,][]{gerin2009}, and toward an interior `core' position \citep[defined by the \ce{DCO+} peak,][]{pety2007}. Of interest to this study, \cite{gratier2013} found that the \chtcn emission is $\sim40$ times brighter at the PDR position than in the `Core'. Similar to protoplanetary disks, this excess in \chtcn could not be explained by gas-phase chemistry alone, and \cite{gratier2013} instead suggested that a combination of UV-mediated surface chemistry with surface desorption processes were responsible. However, models developed by \cite{legal2017}, coupling the \texttt{Meudon PDR} \citep{lebourlot1993,lepetit2006,lebourlot2012} and the \texttt{Nautilus} \citep{hersant2009,ruaud2016} astrochemical codes, could not reproduce the abundance of \chtcn at the PDR position by about two orders of magnitude when taking these processes into account and advanced alternative explanations that {\it(i)} either \chtcn originates from deeper inside of the cloud than previously assumed; or, {\it(ii)} the photo-desorption rate is higher and ice photolysis rate lower than those currently implemented in models; or, {\it(iii)} critical chemical formation pathways are missing in current astrochemical networks.

Another possible explanation for these high nitrile abundances could be the elemental gas-phase C/O ratio. Indeed, the relative elemental gas-phase abundances of oxygen and carbon are known to strongly impact the chemistry of star-forming regions \citep{vandishoeck1998}. For instance, small hydrocarbons, such as \ce{C2H}, \ce{C3H}, \ce{C3H2} and \ce{C4H}, observed in a wide variety of astrophysical objects including PDRs \citep{fuente2003,pety2005,pety2012,cuadrado2015,guzman2015} and protoplanetary disks \citep{dutrey1997,fuente2010,henning2010,qi2013,kastner2015,guilloteau2016,bergin2016,kastner2018,cleeves2018,bergner2019,loomis2019}, are believed to be mainly formed from atomic carbon (i.e. \ce{C+} and/or C). However, atomic carbon is readily converted into CO. Therefore, depending upon the UV-shielding and C/O ratio, more or less carbon can be locked into CO, hampering the production of hydrocarbons and more complex carbon-containing molecules such as \chtcn and \hctn.
In planet-forming disks, a super-solar C/O ratio ($\gtrsim 0.8$) explains the hydrocarbon observations well and is reasonably justified by oxygen removal through water formation and other non-volatile O-bearing species \citep[e.g.][]{hogerheijde2011,cleeves2018}.

Here, we explore whether the observed \chtcn and other complex nitriles in the Horsehead PDR can be explained by pure gas-phase chemistry when taking into account a revised understanding of the cosmic-ray (CR) ionization rate, a more complex gas-phase chemistry network, and most importantly, a C-rich environment. We then carry out a smaller study of complex nitrile production in planet-forming disks with elevated C/O ratios. In \S~\ref{sec:model}, we describe the physical and chemical properties we used and developed within the \texttt{Meudon PDR Code} as well as our fiducial protoplanetary disk model. The resulting molecular abundances and their dependence upon the CR ionization rate, C/O ratio, and complex nitrile formation pathways are presented in \S~\ref{sec:results}. In \S~\ref{sec:discussion}, we discuss the dominant reaction pathways for the four nitrile molecules detected toward the Horsehead nebula - \ctn, \hctn, \chtcn and \ce{CH3NC} - as well as which parameters affect these nitrile abundances.  
Our conclusions are summarized in \S~\ref{sec:conclusion}.

\section{Modeling}
\label{sec:model}

For the PDR chemical investigations we use the \texttt{Meudon PDR Code}, tuned to the physical conditions of the Horsehead nebula, and extended to incorporate a more complete gas-phase chemical network for nitriles up to \chtcn and \ce{CH3NC} in complexity. In the second, smaller part of this paper we use a fiducial protoplanetary disk model previously described in \cite{legal2019} to test whether our nitrile-optimized PDR chemistry can also explain high abundance of complex nitriles in disks.

\subsection{PDR Physical structure}
\label{subsec:phys_struct_meudon}

The \texttt{Meudon PDR Code} is a 1D astrochemical modeling code which considers a stationary plane-parallel slab of gas and dust illuminated by a radiation field \citep{lepetit2006}, which can be introduced at will. Assuming a cloud at steady-state, it solves the physical and chemical conditions at different visual extinction throughout the cloud, taking into account radiative transfer from UV absorption, cooling emissions, and heating processes.

\begin{figure}
    \includegraphics[width=0.48\textwidth]{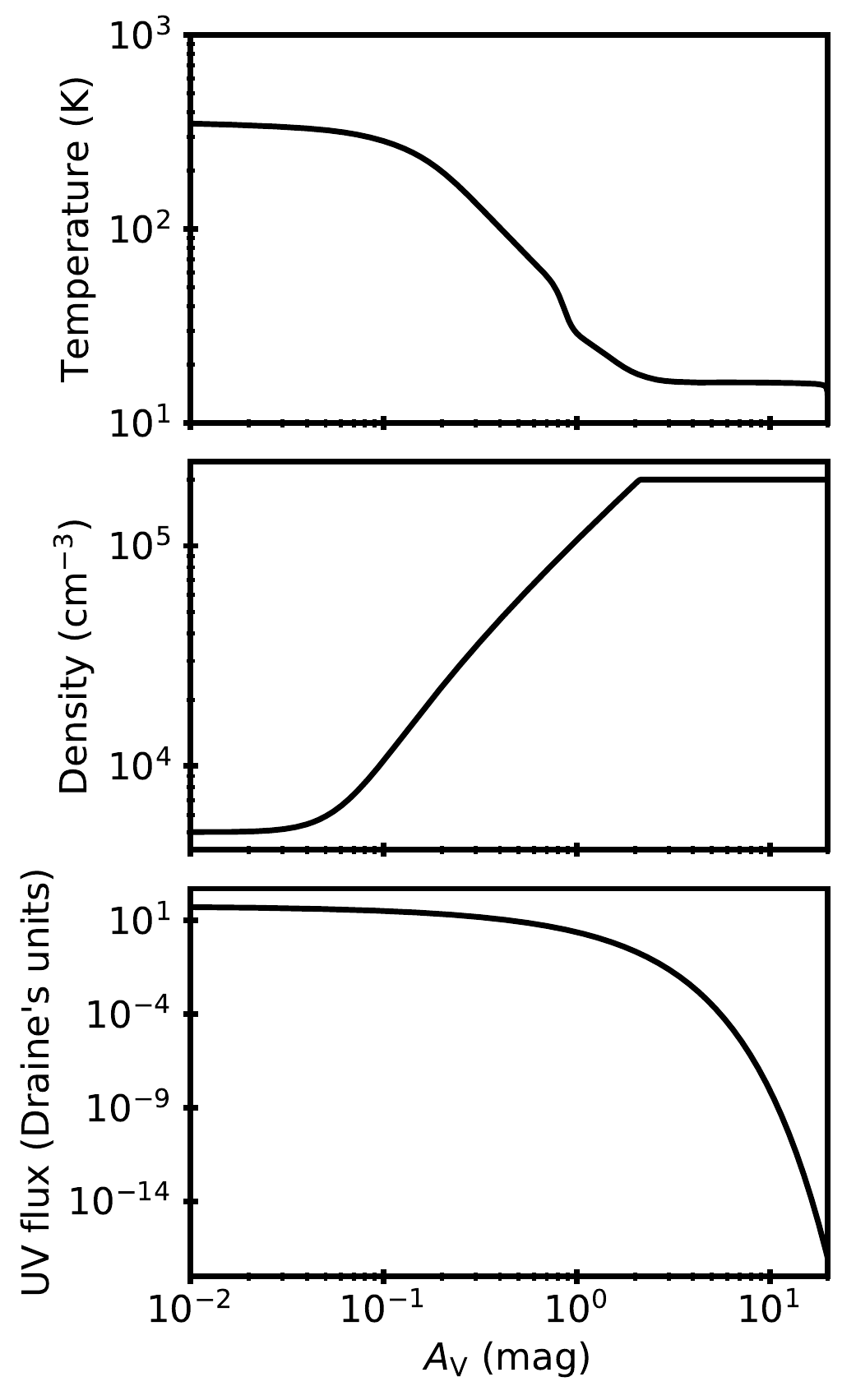}
    \caption{Horsehead nebula profiles of the temperature (top panel), the density (middle panel) and the UV flux (bottom panel) as function of the visual extinction, \av.}
    \label{fig:HH_phys_structure}
\end{figure}

Figure~\ref{fig:HH_phys_structure} displays the typical physical structure we computed for the present study, assuming that the cloud has a fixed pressure of $4 \times 10^6$ K cm$^{-3}$ in the PDR region, and a constant density of $2 \times 10^5$ \ccc in the core \citep{habart2005}, i.e. in our model for \av$\gtrsim 2$~mag. The incident radiation upon this cloud is that of $\sigma$~Ori, an O 9.5 V star system, which results in an incident FUV intensity upon the cloud of about $\chi=60$ \citep[i.e $60~\times$ the ISRF in Draine's units $\approx 60 \times 2.7 \times 10^{-3}$ erg s$^{-1}$ cm$^{-2}$, ][]{draine1978,habart2005}. 
The physical structure shown here was built considering standard initial gas-phase elemental abundances, see Table~\ref{tab:abun}, i.e. a C/O ratio of $\approx 0.46$ \citep{pety2005,goicoechea2006}.

Another parameter to consider is the CR ionization rate $\zeta$. Low energy cosmic rays \citep[10-100 MeV, e.g][]{grenier2015} can penetrate deep into dense clouds, producing ions that drive the gas-phase chemistry via fast ion-neutral reactions. Diffuse clouds usually present higher values of $\zeta$ than denser clouds \citep[e.g.][]{indriolo2015,lepetit2016}, with $\zeta~\approx~(1-5) \times 10^{-17}$ s$^{-1}$ per \ce{H2} being a typical dense cloud value \citep[e.g.][]{goicoechea2009}. We highlight here that what we labeled as $\zeta$ in the present study is the CR ionization rate per \ce{H2}, which corresponds to approximately twice the value of the CR ionization rate per H atom \citep{glassgold1974}. In a previous modeling study of the Horsehead nebula, \cite{rimmer2012} found that chemical predictions are in better agreement with observations when $\zeta$ is allowed to vary across the cloud, considering the following equation adapted from \cite{nath1994} by \cite{rimmer2012}:

\begin{equation}
\zeta = 3.05 \times 10^{-16} (A_{\rm{V}})^{-0.6} + 10^{-17}\s \mathrm{per} \, \ce{H2}.
\label{eq:var-zeta}
\end{equation}

For the PDR position, where \av$\approx 2$~mag, Eq.~(\ref{eq:var-zeta}) gives $\zeta \approx 2 \times 10^{-16}\s$ per \ce{H2}. In \S~\ref{sec:results}, we test the impact of this higher value of $\zeta$ on the nitrile chemistry, by comparison to the canonical value of $\zeta=5\times10^{-17}\s$ per \ce{H2} used in \cite{pety2005} and \cite{goicoechea2006}.

Lastly, while the public version of the \texttt{Meudon PDR Code (v.1.5.2)} does not include grain chemistry, it does model the formation of H$_2$ on grains and computes the charge and temperature distribution of grains. In this study, we kept the default grain size distribution, i.e with grain radius from $1 \times 10^{-3}$ to 0.3~$\mu$m and their relative abundances described by the MRN distribution \citep{mathis1977}.

\subsection{PDR model Chemistry}
\label{subsec:chem_pres}

\begin{table}
\begin{center}
\caption{Initial gas-phase elemental abundances}
\label{tab:abun}
\begin{tabular}{cc}
		\hline
        \hline
        Species & $n_i/n_{\text{H}}$  $^{(a)}$ \\
        \hline
        He      & 0.1 \\
        O$^{(b)}$       & $3.02\times10^{-4}$\\
        C   & $1.38\times10^{-4}$ \\
        N       & $7.95\times10^{-5}$ \\
        S   & $3.50\times10^{-6}$ \\
        Si & $1.73\times10^{-8}$ \\
        Fe  & $1.70\times10^{-9}$ \\
        \hline
        \hline
\end{tabular}
\end{center}
\tablecomments{${(a)}$ from \cite{pety2005} \& \cite{goicoechea2006}
${(b)}$ To test the impact of the C/O ratio, we varied the oxygen elemental abundance in the range $[3.45-0.92]\times10^{-4}$ (see \S~\ref{subsec:C_O_impact}).}  
\end{table}

Each model was performed using the same initial abundance set as in \cite{pety2005} and \cite{goicoechea2006}, except for the oxygen abundance that we varied in some models to explore the impact of the C/O ratio on the chemistry (Table~\ref{tab:abun}).
We updated and extended the \texttt{PDR Meudon Code  (v.1.5.2)} chemical network with 39 species and 913 reactions relevant to the chemistry of \ctn, \hctn, and \chtcn, that we extracted from the KIDA database\footnote{http://kida.obs.u-bordeaux1.fr/} for most of them. We also extended the chemical network to the chemistry of \chtnc, based on theoretical studies \citep[e.g.][]{defrees1985} and the chemistry of its isomer \chtcn (see \S~\ref{subsec:ch3cn_ch3nc}).
In total, our network is composed of 191 species and 3616 chemical reactions, including gas-phase bi-molecular reactions (i.e. radiative associations, ion-neutral and neutral-neutral reactions), recombinations with electrons, ionization and dissociation reactions by direct cosmic rays and secondary photons (i.e. photons induced by cosmic rays), and by UV-photons \citep[see ][for rate formulae details]{lepetit2006}. The critical reactions discussed in this paper are summarized in Table~\ref{tab:new_reactions}, with rates and references.

UV-photo-reactions are expected to play a crucial role in PDR chemistry. The \texttt{Meudon PDR Code} allows the choice between two different methods to compute the photo-reaction rates: 1) if the photo-ionization and/or photo-dissociation cross-sections of the molecule is known, the most accurate approach consists in integrating this cross-section over the radiation field at each given position in the cloud; 2) if the photo-cross-section of the molecule is unknown, an analytical expression as function of the visual extinction is estimated, i.e. considering fixed fitted parameters for each molecules and computing their photo-rates as function of the visual extinction \cite[see for an example Eq.~(14) of ][]{heays2017}. 
We updated the cross-sections of all the molecules available in the Leiden database\footnote{https://home.strw.leidenuniv.nl/$\sim$ewine/photo/} that are included in our chemical network in the \texttt{Meudon PDR Code} \citep{heays2017}. 

\subsection{Protoplanetary disk physical structure} 
\label{subsec:disk_model_phys_struct}
\begin{figure*}
    \centering
    \includegraphics[scale=0.43]{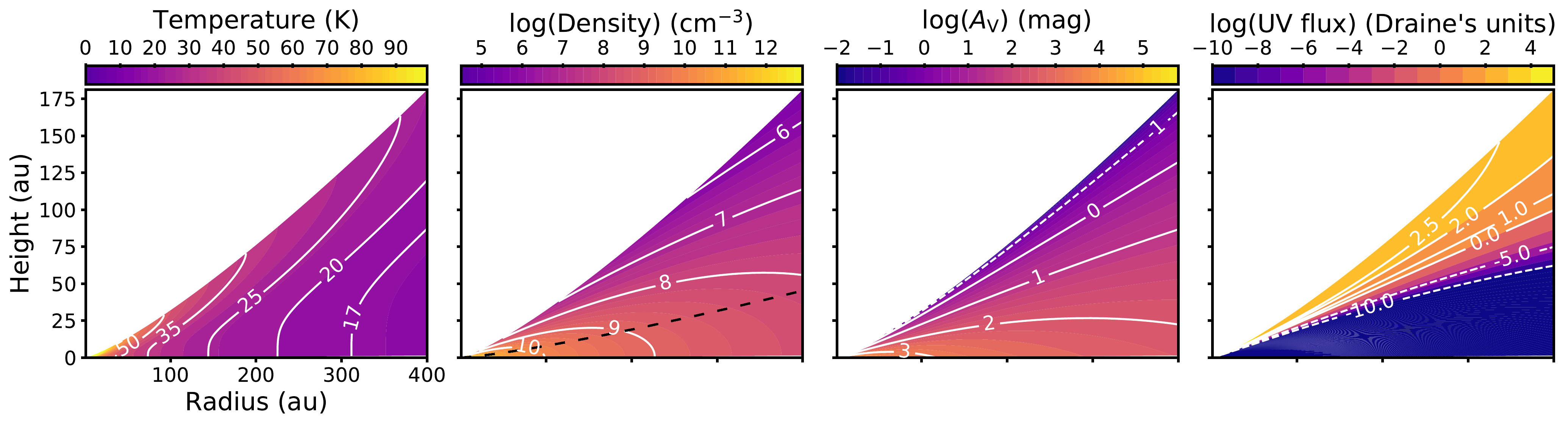}
    \caption{Disk physical structure fed in our fiducial protoplanetary disk astrochemical model. The 2D temperature ({\it first panel}), density ({\it second panel}), visual extinction ({\it third panel}) and UV flux ({\it fourth panel}) profiles are represented as functions of disk radius versus height, both in au. The dashed black line, on the densities panels, delineates 1 scale height.}
    \label{fig:disk_phys_struct}
\end{figure*}

Our fiducial protoplanetary disk astrochemical model is based on the MWC~480 disk model of \cite{legal2019}, which consists in a 2D parametric physical structure onto which the chemistry is post-processed (see \S \ref{subsec:disk_model_chem}). The disk physical structure assumes a disk that is symmetric azimuthally and with respect to the midplane. Thus, it can be described in cylindrical coordinates centered on the inner star along two perpendicular axes characterizing the radius and height in the disk. Figure~\ref{fig:disk_phys_struct} represents the profiles of the gas temperature, density, visual extinction and UV flux throughout the disk, for which the parameterization is briefly summarized below and further described in \cite{legal2019}.

For a given radius $r$ from the central star, the vertical temperature profile is computed following the formalism of \cite{rosenfeld2013} and \cite{williams2014}, originally developed by \cite{dartois2003}:

\begin{equation}
\small
T(z) = \left\{
    \begin{array}{ll}
        T_{\rm{mid}}+(T_{\rm{atm}}-T_{\rm{mid}})\left[ \sin \left(\frac{\pi z}{2z_q}\right) \right]^{2\delta}&\mbox{if} \, z<z_q\\
        T_{\rm{atm}}&\mbox{if} \, z\ge z_q,
    \end{array}
\right.
\label{eq:ture}
\end{equation}
where $T_{\mathrm{mid}}$ and $T_{\mathrm{atm}}$ are respectively the midplane and atmosphere temperatures that
vary as power law of the radii \citep{beckwith1990,pietu2007,legal2019}. $z_q=4H$ with $H$ the pressure scale height that, assuming vertical static equilibrium, can be expressed as follows:

\begin{equation}
H=\sqrt \frac{k_{\rm{B}} \, T_{\rm{mid}} \,r^3}{\mu \,m_{\rm{H}}\, G \,M_\star},
\end{equation}
with $k_{\rm{B}}$ the Boltzmann constant, $\mu=2.4$ the reduced mass of the gas, $m_{\rm{H}}$ the proton mass, $G$ the gravitational constant, and $M_\star$ the mass of the central star. The midplane temperature $T_{\mathrm{mid}}$ is estimated following a simple irradiated passive flared disk approximation \citep[e.g.][]{chiang1997,dullemond2001,huang2018}:
\begin{equation}
    T_{\rm{mid}}(r)\approx \left(\frac{\varphi L_\star}{8\pi r^2 \sigma_{\rm{SB}}}\right)^{1/4},
    \label{eq:Tmid_Rc}
\end{equation}
with $L_\star=24~L_\odot$ the stellar luminosity \citep{andrews2013}, $\sigma_{\rm{SB}}$ the Stefan-Boltzman constant and $\varphi=0.05$ a typical flaring angle. The atmosphere temperature, $T_{\mathrm{atm}}$, is based on observational constraints. So here we consider $T_{\mathrm{atm}}=T_{\mathrm{atm},100\,\rm{au}}(\frac{r}{100\,\rm{au}})$, with $T_{\mathrm{atm},100\,\rm{au}}$=48~K from \cite{guilloteau2011}.

The disk is assumed to be in hydrostatic equilibrium. Thus, for a given vertical temperature profile, the vertical density structure is determined by solving the equation of hydrostatic equilibrium, as described from Eq. (17) to (20) in \cite{legal2019}.
The surface density of the disk is assumed to follow a simple power law varying as $r^{-3/2}$ \citep{shakura1973,hersant2009}: 
\begin{equation}
\Sigma (r) = \Sigma_{R_c} \left( \frac{r}{R_c}\right)^{-3/2},
\end{equation}
where $\Sigma_{R_c}$ is the surface density at the characteristic radius that can be expressed as function of the mass of the disk, $M_{\rm{disk}}$, and its outer radius, $R_{\rm{out}}$:
\begin{equation}
\Sigma_{R_c}=\frac{M_{\rm{disk}} R_c^{-3/2}}{4\pi \sqrt{R_{\rm{out}}}},
\end{equation}
with here $M_{\rm{disk}}=0.18~M_\odot$ \citep{guilloteau2011}.

The visual extinction profile is derived from the hydrostatic density profile using the gas-to-extinction ratio of $N_{\ce{H}}/A_{\rm{V}}=1.6\times10^{21}$ \citep{wagenblast1989}, with $N_{\ce{H}}=N(\ce{H})+2N(\ce{H2})$ the vertical hydrogen column density of hydrogen nuclei. This gas-to-extinction ratio assumes a typical mean grain radius size of 0.1~$\mu$m and dust-to-mass ratio of 0.01, consistent with model assumptions.

Finally, the UV flux profile is computed considering the UV flux impinging the disk convolved with the visual extinction profile. 
The unattenuated UV flux factor, $f_{\rm{UV}}$, at a given radius $r$ depends on both the photons coming directly from the central embedded star and on the photons that are downward-scattered by small grains in the upper atmosphere of the disk. Thus, following \cite{wakelam2016}, we consider:
\begin{equation}
f_{\rm{UV}}=\frac{f_{\rm{UV},R_c}/2}{\left(\frac{r}{R_c}\right)^2+\left(\frac{4\rm{H}}{R_c}\right)^2}.    
\end{equation}

\subsection{Protoplanetary disk chemical model} 
\label{subsec:disk_model_chem}

The disk chemistry is computed time-dependently in 1+1D using the gas-grain astrochemical model \texttt{Nautilus (v.1.1)} \citep{hersant2009,wakelam2016} in three phase mode \citep{ruaud2016}, i.e. including gas-phase, grain-surface and grain-bulk chemistry \citep[see ][for more details]{legal2019}.
First, the chemical evolution of a representative starless dense molecular cloud is modeled up to a characteristic age of $1\times10^{6}$ years \citep[e.g.][]{emelgreen2000,hartmann2001}. 
For this 0D model we use typical constant physical conditions: grain and gas temperatures of 10~K, a gas density of $2\times 10^{4} \ccc$ and $\zeta=5\times 10^{-17}\s$ per \ce{H2}; this parent molecular cloud is also considered to be shielded from external UV photons by a visual extinction of 30 mag. For consistency, we use the same initial abundances as for our PDR model (see  Table~\ref{tab:abun}) for this first simulation step. The outcoming chemical gas and ice compositions of this parent molecular cloud are then used as initial chemistry for our 1+1D disk model. Second, we ran the chemistry of our 1+1D disk model up to one million years, typical chemical age of a disk when grain growth is not considered \citep[e.g.][]{cleeves2015}. While the disk chemistry has not reached steady state at that time, its evolution is slow enough that the results presented here hold for a disk twice younger or older. Note that in contrast to the PDR model, the disk chemical code does include grain surface reactions. However the grain-surface reactions pathways to CH$_3$CN and HC$_3$N, the two molecules or particular interest for this study, remain poorly constrained. 

\section{Results}
\label{sec:results}

\subsection{Impact of cosmic-ray treatment}
\label{subsec:res_new_chem_and_zeta}

\begin{figure}
    \centering
    \includegraphics[width=.47\textwidth]{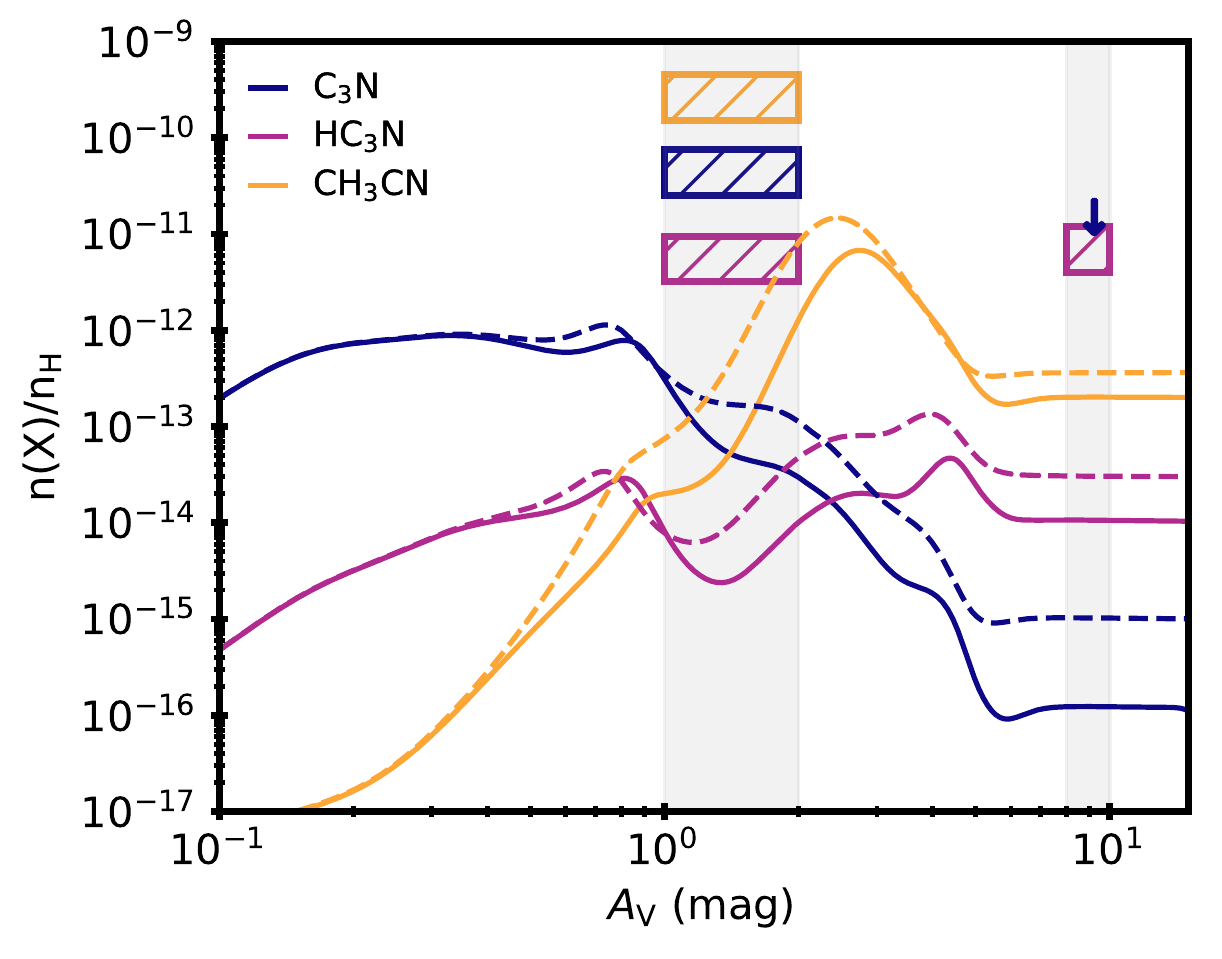}
    \caption{Computed \ctn (dark blue) \hctn (purple) and \chtcn (orange) abundances with respect to H nuclei, as function of the visual extinction \av obtained with the {\it standard} model (solid lines) and with the {\it high-$\zeta$} model (dashed lines), see \S~\ref{subsec:res_new_chem_and_zeta}. These model results are compared to the observations from \cite{gratier2013} (dashed boxes and arrow). 50\% error bars are included on the observations.  The PDR (1~mag $<$ \av $<$ 2~mag) and Core (\av $\gtrsim$ 8~mag) regions are shaded.}
    \label{fig:meudon_goicoechea06_ini}
\end{figure}

Figure~\ref{fig:meudon_goicoechea06_ini} presents the abundances of \ctn, \hctn and \chtcn computed with the \texttt{Meudon PDR Code} as function of the visual extinction \av, for two models. 
Both models consider our new chemical network and the initial gas-phase elemental abundances prescribed in \cite{pety2005} and \cite{goicoechea2006} (see Table~\ref{tab:abun}) but each model uses a different CR ionization rate. The {\it standard} model uses the CR ionization rate canonical value of $\zeta=5\times10^{-17}\s$~per \ce{H2} \citep{pety2005,goicoechea2006}, and the {\it high-$\zeta$} model uses a higher CR ionization rate of $\zeta=2\times10^{-16}\s$ per \ce{H2}, as calculated from (Eq.~\ref{eq:var-zeta}). 
By impacting the ion abundances in molecular clouds, the CR ionization rate indirectly drives the abundances of their daughter neutral molecules (see \S~\ref{subsec:phys_struct_meudon}). The nitrile abundances are indeed higher with the {\it high-$\zeta$} model than with the {\it standard} model, but both models under-predict by several orders of magnitude the abundances observed toward the Horsehead nebula.

\cite{rimmer2012} showed that a varying $\zeta$ across the cloud tends to produce more accurate results. However, for the visual extinctions associated with the PDR region and for molecules of interest studied here, varying $\zeta$ across the cloud does not significantly impact the results compared to a constant-$\zeta$ model. In the Core region, the varying-versus-constant $\zeta$ model abundances are about half an order of magnitude different.

Changing the CR ionization rate also impacts the gas temperature, as shown in the bottom panels of Fig.~\ref{fig:meudon_co}. Typically, increasing $\zeta$ shifts the temperature gradient closer to the PDR's edge. As a result, a higher $\zeta$ leading to higher temperatures and thus higher reaction rates, the absolute abundances of nitriles are slightly increased in the PDR with a {\it high-$\zeta$} model. In addition, likely due to the shift in temperature, the abundance patterns are shifted outwards (i.e. closer to the illuminated edge of the PDR) in the case of a higher $\zeta$.
In summary, while the CR treatment has an impact on the complex nitrile chemistry, increasing nitrile abundances by almost an order of magnitude in the PDR region, the effect is small compared to the mismatch between models and observations (see Fig.~\ref{fig:meudon_goicoechea06_ini}).

\subsection{Impact of the C/O ratio}
\label{subsec:C_O_impact}

Major carriers of O and C are expected to freeze out under different conditions. In particular a substantial amount of O can become incorporated into water ice which is one of the least volatile common interstellar molecules, resulting in an elevated C/O ratio in the gas-phase. In order to mimic the differential freeze-out of volatiles on grains, we varied the oxygen gas-phase elemental abundance from $3.45\times10^{-4}$ to $9.2\times10^{-5}$ while keeping the carbon abundance fixed. This led to a variation of the C/O ratio from 0.4 to 1.5. The lowest considered O abundance is a factor of two higher than the CO abundance derived in the Horsehead PDR \citep[$5.6\times10^{-5}$,][]{pety2005}, while the highest considered O abundances is below the cosmic O abundance of $4.9\times10^{-4}$ \citep{asplund2009} to $5.75\times10^{-4}$ \citep{przybilla2008}. 
The choice of fixing the carbon elemental abundance and varying the oxygen one, is also justified by the fact that, between these two elements, the elemental gas-phase abundance of oxygen is the less constrained \citep{jenkins2009,whittet2010,jones2019}, as discussed in \cite{legal2014}. 

\begin{figure*}
    \centering
    \includegraphics[width=0.85\textwidth]{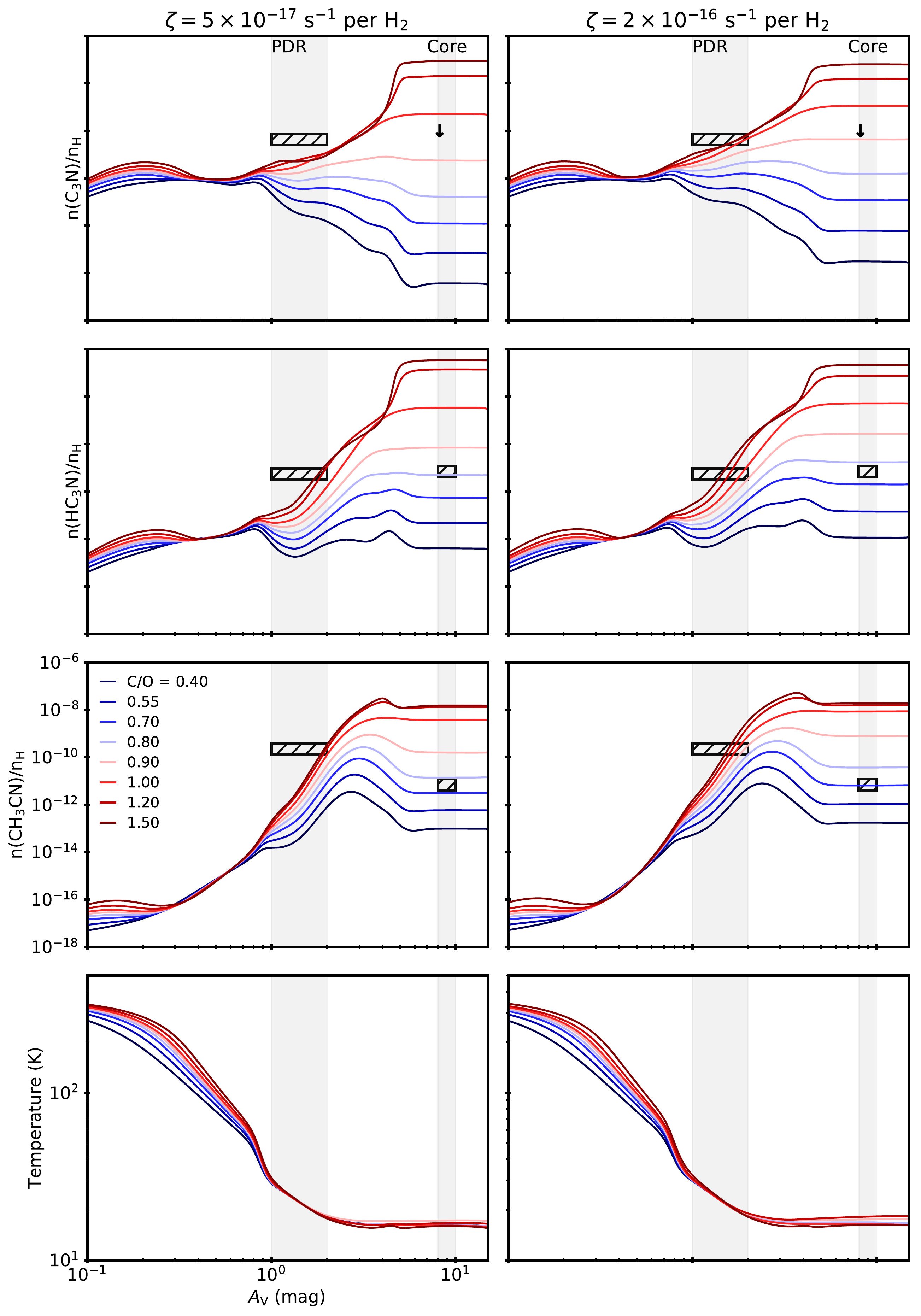}
    \caption{\ctn, \hctn and \chtcn abundances with respect to H nuclei, as well as the gas temperature, computed with our model of the Horsehead nebula as function of the visual extinction, \av, for varying C/O and the standard- (left column) and high- (right column) CR ionization rates considered in this study. The observations from \cite{gratier2013} are represented by the black hatched boxes, which consider 50\% error bars, and the downward arrow. The core (\av $>$ 8~mag) and PDR (1~mag $<$ \av $<$ 2~mag) regions are shaded.}
    \label{fig:meudon_co}
\end{figure*}

Figure~\ref{fig:meudon_co} shows the impact of the gas-phase C/O ratio on the abundances of \ctn, \hctn and \chtcn and on the gas temperature as function of the visual extinction \av in our model of the Horsehead nebula. An O-poor chemistry (i.e. a high C/O ratio) results in higher abundances of the three nitriles.  For a CR ionization rate of $\zeta=2\times10^{-16}\s$ per \ce{H2} and a gas-phase C/O ratio in the range $0.9-1.5$, our new gas-phase chemistry model can reproduce the three nitrile observations at the PDR position within an order of magnitude. As for the Core position, our best fit models are found for lower C/O ratios, in the range $0.6-0.9$. This lowering of C/O with increasing visual extinction could be explained by photon-mediated release of refractory carbon into gas phase in the PDR region, and/or the onset of freeze out of carbon species in the core region. Whatever the mechanism, the decrease of C/O with increasing visual extinction suggests that the gas-phase C/O ratio vary across astrophysical objects.

It is also important to mention that the chemical rates used in astrochemical models sometimes present large uncertainties. We ran two additional models to test the impact of such uncertainties on the major reaction rates listed in Table~\ref{tab:new_reactions} which are driving the complex nitrile chemistry. These additional simulations compute the chemistry with {\it(i)} the maximum allowed rates, and {\it(ii)} the minimum allowed rates. The results are that the nitrile abundances of interest for this study vary by less than a factor of three in the PDR and Core regions, which is small compared to the more than two orders of magnitude mismatch between observations and models using the standard C/O value. Though this does not constitute a rigorous detailed sensitivity analysis such as those developed for instance by \cite{vasyunin2004,vasyunin2008} and \cite{wakelam2005,wakelam2006,wakelam2010}, our simple analysis suggests that our results are robust.

Since we are mainly interested in the PDR nitrile chemistry, we consider our best-fit model the model with the lowest C/O ratio that reproduces at the PDR position the three complex nitrile abundances shown Fig.~\ref{fig:meudon_co}. The model with $\zeta=2\times10^{-16}\s$ per \ce{H2} and C/O $=1.0$ fulfills these criteria.  

\begin{figure*}
    \centering
    \includegraphics[width=1.0\textwidth]{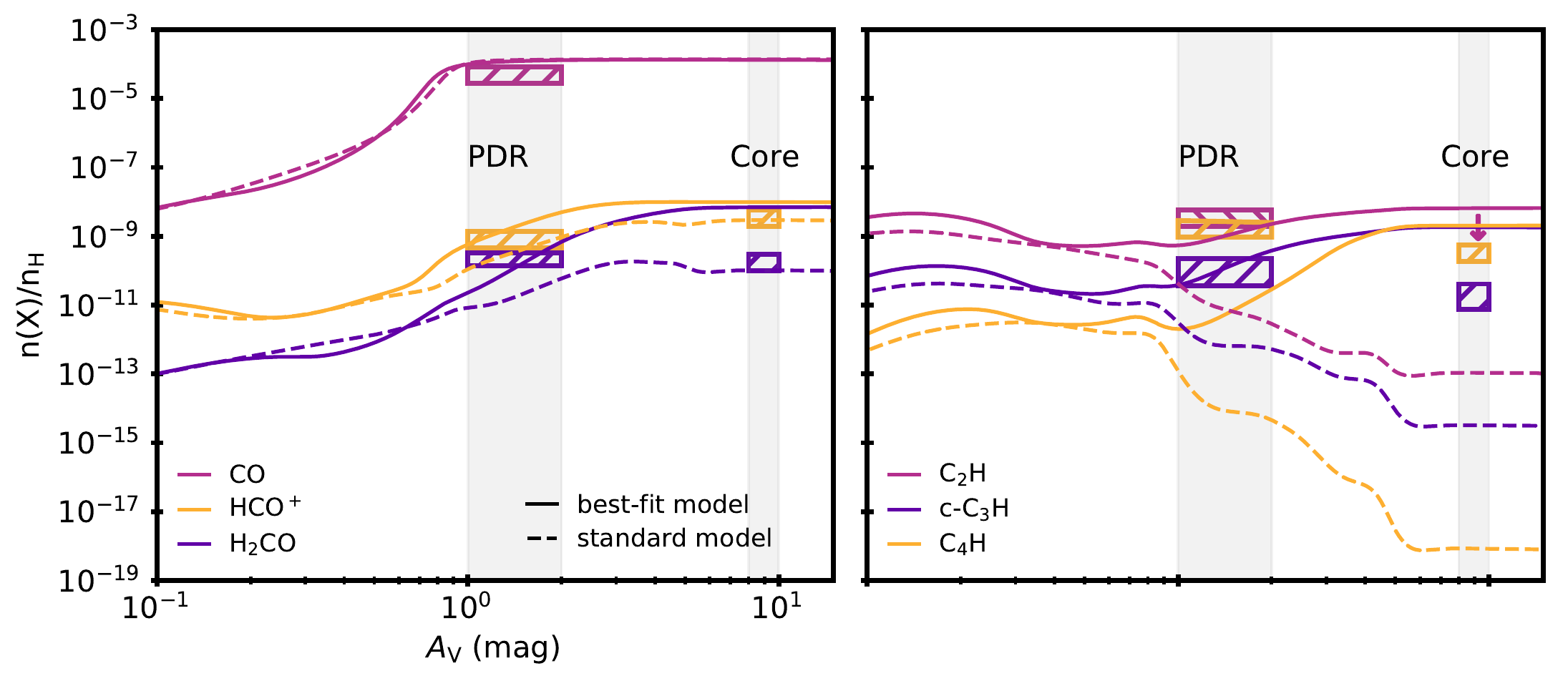}
    \caption{CO, \ce{HCO+}, \ce{H2CO}, \ce{C2H}, \ce{c-C3H} and \ce{C4H} abundances with respect to H nuclei, as function of the visual extinction \av computed with our best-fit model (solid lines) and standard model (dashed lines) of the Horsehead nebula, compared to published observations \citep{pety2005,goicoechea2009,guzman2011,guzman2015} represented by the filled and hatched boxes and the downward arrows. 50\% error bars are included on the observations.  The core (\av $>$ 8~mag) and PDR (1~mag $<$ \av $<$ 2~mag) regions are shaded.}
    \label{fig:other_than_cyanides}
\end{figure*}

Figure~\ref{fig:other_than_cyanides} shows the modeled abundances, obtained with our best-fit model, of other typical oxygenated and carbonated molecules that were also observed toward the Horsehead nebula, i.e CO \citep{pety2005}, 
HCO$^+$ \citep{goicoechea2009}, \ce{H2CO} \citep{guzman2011}, and the hydrocarbons C$_2$H, \ce{c-C3H}, and \ce{C4H} \citep{pety2005,guzman2015}. For comparison, the {\it standard} model results are also presented in the same figure.
The C/O variation does not significantly impact the CO abundance throughout the cloud, which is consistent with the fact that CO is the main reservoir of carbon and the carbon elemental abundance is fixed in our models. Our best fit model displays better agreement between model and observations in the PDR for the other O-bearing molecules we consider, \ce{H2CO} and \ce{HCO+}.

With regards to the hydrocarbons, our best model better matches the observations at the PDR position, generally by orders of magnitude, compared to the standard model. This is an expected results because atomic O is a main destroyer for small hydrocarbons such as \ce{C2H}, \ce{c-C3H} and \ce{C4H} \citep{millar1987,millar1990}. Thus, while C is kept constant, diminishing O increases the hydrocarbon abundances. However, even our best fit model does not fully reproduce the observed abundances (see for instance \ce{C4H} in Fig.~\ref{fig:other_than_cyanides}), indicative of that the C/O ratio does not provide a complete explanation for the abundant hydrocarbon chemistry in the Horsehead PDR. 

Regarding the Core position, our best fit model generally over-predicts the observations, but this might simply be explained by the fact that freeze-out on grains is not included in our model.
It could also be that the gas-phase C/O ratio decreases within the nebula, as suggested by the C/O grid results shown in Fig.~\ref{fig:meudon_co} when compared to the observations in each observed positions.

\subsection{\ce{CH3NC} vs \ce{CH$_3$CN} chemical pathways}
\label{subsec:ch3cn_ch3nc}

\begin{figure}
    \centering
    \includegraphics[width=0.48\textwidth]{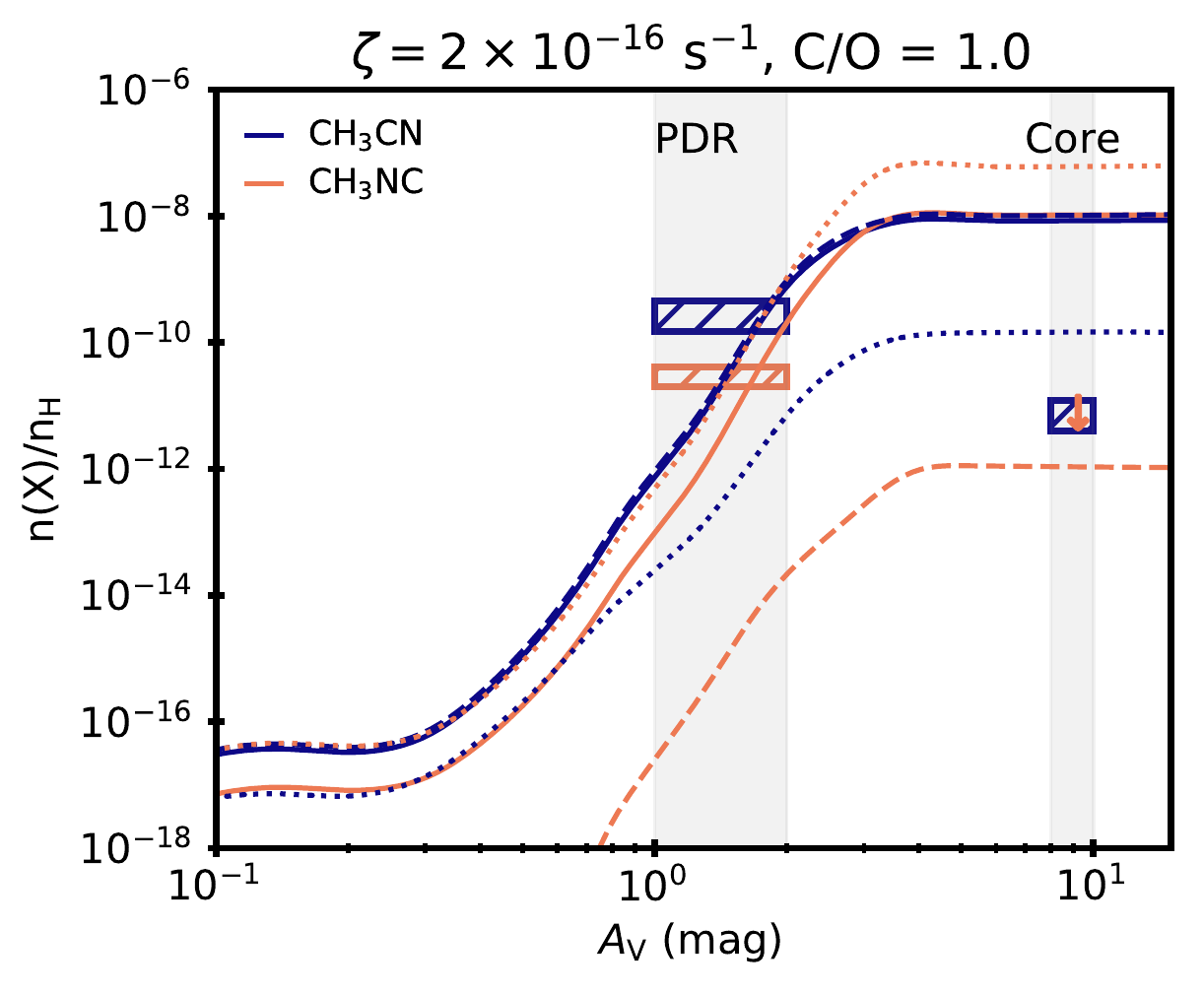}
    \caption{\ce{CH3CN} (dark blue) and \ce{CH3NC} (orange) abundances with respect to H nuclei, as function of the visual extinction, \av, in the modeled Horsehead nebula. Three different models, based on our best-fit model (see \S~\ref{subsec:C_O_impact}), are depicted here, testing the izomerization branching ratio of the reaction pathway (\ref{eq:2}): {\it(i)} 100\% (dotted lines); {\it(iii)} 80\% (solid lines - best model); {\it(iii)} 0\% (dashed lines).  The model results are also compared to published observations from \cite{gratier2013}, where 50\% error bars are included (dashed boxes and downward arrow). The core (\av $>$ 8~mag) and PDR (1~mag $<$ \av $<$ 2~mag) regions are shaded.}
    \label{fig:ch3cn_ch3nc}
\end{figure}

Another interesting nitrile molecule to study is the methyl isocyanide (\ce{CH3NC}), the isomer of methyl cyanide (\ce{CH3CN}). First detected toward Sgr B2 \citep{cernicharo1988,remijan2005}, \ce{CH3NC} was also detected toward the Horsehead nebula \citep{gratier2013}, Orion  KL \citep{lopez2014}, and more recently toward the solar-type binary protostar IRAS~16293-2422 \citep{calcutt2018}. A few theoretical and experimental studies have investigated the isomers' chemistry and their abundance ratio \citep{huntress1979,defrees1985,anicich1995}, and converged on the same major gas-phase production pathways for both via the reaction:
\begin{align}
\mhchemoptions{arrow-min-length=1em}
\ce{ CH3+ + HCN  -> (CH3NCH+)$^{*}$ ->[$k_2$] CH3NCH+ + $h\nu$},\label{eq:1}\\
\mhchemoptions{arrow-min-length=1em}
\ce{->[$k_3$] CH3CNH+ + $h\nu$}\label{eq:2},
\end{align}
with $k_2$ and $k_3$ given in Table~\ref{tab:new_reactions}, followed by the dissociative recombinations of both protonated ions \ce{CH3NCH+} and its isomer \ce{CH3CNH+} to form \ce{CH3NC} and \ce{CH3CN}, respectively (see reaction rates in Table~\ref{tab:new_reactions}). However, the branching ratio is poorly constrained and depends on the stabilization processes of the intermediate complex (\ce{CH3NCH+})$^{*}$  \citep[e.g.][]{anicich1995}. 
Due to its lower energy state, \ce{CH3CNH+} is found to be the major product of the reaction \ce{CH3+ + HCN}~(\ref{eq:2}). However, its formation requires the isomerization of the intermediate complex (\ce{CH3NCH+})$^{*}$, which likely happens due to collisions with a third body. Thus, the ratio between the two isomeric ions depends on the competition between the relaxation and isomerization rates of the intermediary complex. The resulting \ce{CH3NCH+}/\ce{CH3CNH+} ratio was estimated to lie in the range 0.1--0.4 by one theoretical study and assumed to propagate to a \ce{CH3NC}/\ce{CH3CN} ratio of 0.1--0.4 via the respective subsequent dissociative recombinations \citep{defrees1985}. 

Here, we investigated the impact of the branching ratios in between the pathways~(\ref{eq:1}) and (\ref{eq:2}) on the resulting \ce{CH3NC}/\ce{CH3CN} ratio in our PDR model. Figure~\ref{fig:ch3cn_ch3nc} presents the results obtained using our best fit model and three different branching ratios leading to: 100\%, 80\% and 0\% of isomerization. The best fit results are obtained for a branching ratio of 80\% (i.e. \ce{CH3NC}/\ce{CH3CN}~$\sim 0.2$), in agreement with the theoretical calculation of \cite{defrees1985}. However, to our knowledge the \ce{CH3NCH+} dissociative recombination has not been studied yet and even though the rate of the \ce{CH3CND+} dissociative recombination was measured \citep{vigren2008} its branching ratio remains uncertain \citep[e.g.][]{plessis2010,plessis2012,loison2014}. It would thus be interesting to study whether the dissociative recombination of \ce{CH3CNH+} and \ce{CH3NCH+} could lead to disproportionate prevalence of each initial isomer. Further theoretical and experimental studies are therefore needed to assess the validity of our astrochemically motivated branching ratios.

\subsection{Complex nitrile production in a protoplanetary disk with a high C/O ratio}

\begin{figure*}
    \centering
    \includegraphics[width=1.0\textwidth]{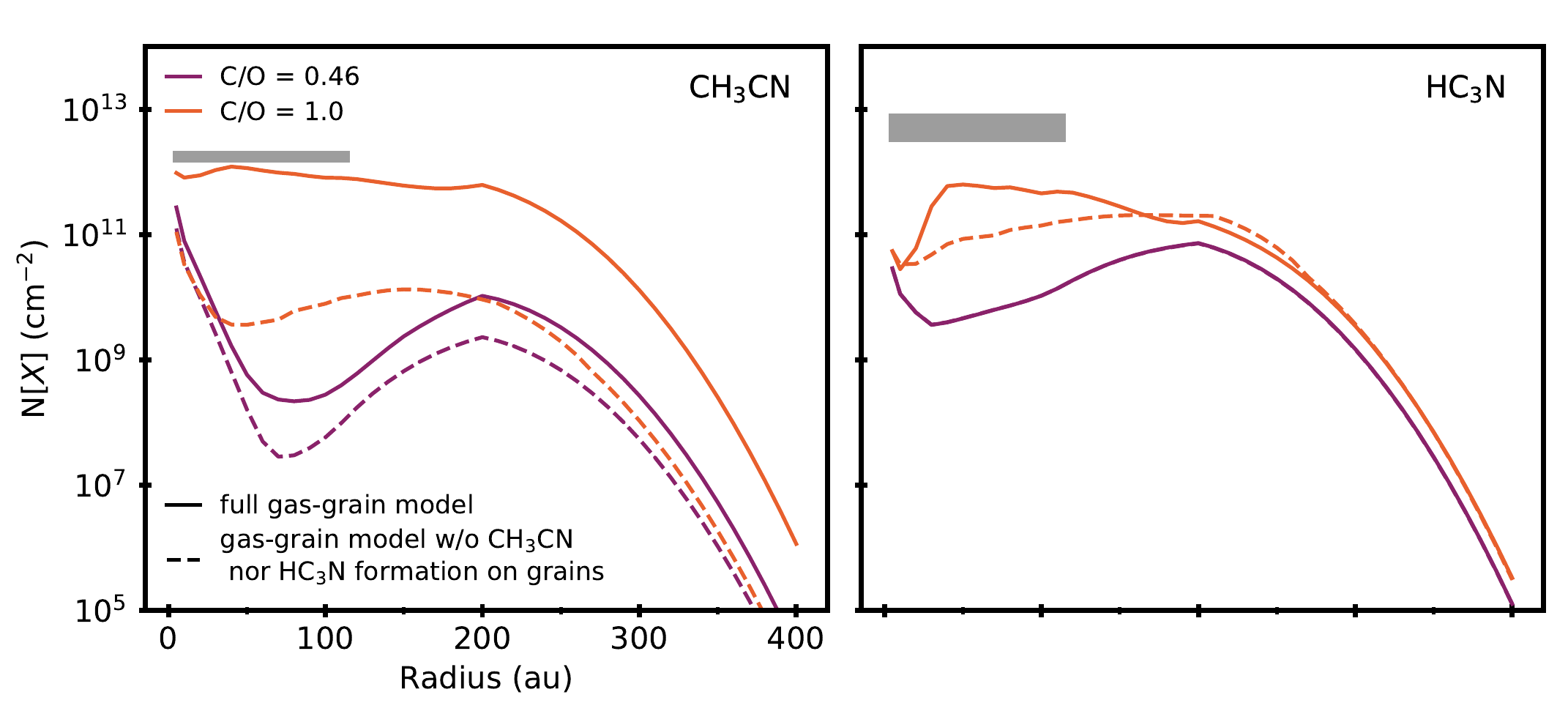}
    \caption{Radial profile of the column density of \chtcn and \hctn computed with our fiducial protoplanetary disk astrochemical model for four different models, differing in {\it (i)} their C/O ratio, with {\it (a)} C/O$ = 0.46$ (in purple) as in our PDR {\it standard} model and {\it (b)} C/O$ = 1.0$ (in orange) as in our PDR best fit model; {\it (ii)} their grain chemical network that includes (solid lines) or not (dashes lines) the formation of \chtcn and \hctn on grains. The horizontal gray lines represent the column density derived from the MWC~480 disk observations of \citep{bergner2018}.\\}
    \label{fig:disk_model_c_o_comp}
\end{figure*}

To test if our new understanding of the complex nitrile PDR chemistry can be generalized to disks, we used a fiducial protoplanetary disk astrochemical model, loosely based on the disk around MWC 480, from \cite{legal2019} described \S~\ref{subsec:disk_model_phys_struct} and \S~\ref{subsec:disk_model_chem}. We ran the chemical post-processing for two different C/O ratios: {\it (i)} C/O$ = 0.46$, as in our PDR {\it standard} model, and {\it (ii)} C/O$ = 1.0$ as in our PDR best fit model. For each of these C/O ratios we ran two disk models, a full gas-grain model and a gas-grain model where \chtcn and \hctn are only formed in the gas phase. The results of this total of four disk models on the abundances of \hctn and \chtcn are shown in Fig.~\ref{fig:disk_model_c_o_comp}. A huge gap is observed from $\sim~25$~to $\sim200$~au in the computed nitrile column densities for the standard C/O ratio disk models, and disappears for higher C/O. The prevalence of grains' pathway formation increases with C/O, and even becomes negligible in the formation of \hctn for standard C/O. Even though our disk model also includes grain chemistry, the main result from our PDR study holds for disk astrochemistry, i.e. that an elevated C/O ratio better reproduce the nitrile observations. Without any tuning of our disk model, our best-fit model predictions are within an order of magnitude for the \chtcn case. For the \hctn, the results are in agreement, at the order of magnitude level, for the inner 100 au of the disk, where likely most of the emission originates \citep{bergner2018}. 

\section{Discussion}
\label{sec:discussion}

\subsection{Nitrile formation pathways}

In our PDR models, the \ctn formation is dominated by the reaction:
\begin{equation}
  \ce{c-C3H + N ->[$k_4$] H + C3N},
  \label{reac:c3n_form}
\end{equation}
with $k_4$ given in Table~\ref{tab:new_reactions}.
\ce{c-C3H} being itself mainly produced by the electronic recombination of \ce{c-C3H2+}. Thus, the under-prediction of \ctn by our model in the PDR region could be explained by the under-prediction of \ce{c-C3H} (see Fig.~\ref{fig:other_than_cyanides}).
Therefore, for a fixed abundance of N, a carbon-enriched medium would enhance the production of \ctn. As for its destruction, photo-dissociation dominates in the PDR (see the corresponding reaction rate Table~\ref{tab:new_reactions}), 
but depending on the C/O ratio two other reactions also participate in the \ctn destruction, i.e.:
\begin{align}
\ce{O + C3N ->[$k_5$] CO + CCN}\label{eq:dest_c3n_by_o},\\
\ce{C + C3N ->[$k_6$] C3 + CN}\label{eq:dest_c3n_by_c},
\end{align}
with $k_5$ and $k_6$ given in Table~\ref{tab:new_reactions}.
Reaction (\ref{eq:dest_c3n_by_o}) even becomes the primary destruction pathway of \ctn in O-rich (C/O~$\approx 0.4$) PDR. While, a priori, it may seem odd for a photo-dissociation process to not be the primary destruction mechanism in PDR, in our model this is explained by the high concentration of atomic oxygen in this region for low C/O ratios. 
In O-poor (C/O~$\gtrsim 1$) PDR, \ctn is primarily destroyed through a combination of photo-dissociation and by atomic carbon~(\ref{eq:dest_c3n_by_c}).

\hctn is formed from a variety of different reactions involving carbon- and nitrogen-containing molecules:
\begin{align}
\ce{HC3NH+ + e- ->[$k_7$] H + HC3N},\label{reac:hc3n_form1}\\
\ce{C4H + N ->[$k_{8}$] C + HC3N},\\
\ce{C + H2CCN ->[$k_{9}$] H + HC3N},\\
\ce{c-C3H2 + N ->[$k_{10}$] H + HC3N},\\
\ce{C2H2 + CN ->[$k_{11}$] H + HC3N},
\end{align}
with $k_7$ to $k_{11}$ given in Table~\ref{tab:new_reactions}.
Thus, with a lower amount of one of the main hydrocarbon destroyer, atomic O, and a higher amount of `free' carbon in the gas phase, more reaction pathways meaningfully contribute to the formation of \hctn, via these diverse carbon-rich intermediates. 
\hctn is mainly destroyed by UV-photons up to an \av$\approx4$~mag, via the reaction:
\begin{equation}
\ce{HC3N + $h\nu$ ->[$k_{12}$] C2H + CN},
\end{equation}
with $k_{12}$ given in Table~\ref{tab:new_reactions}.
For \av$\gtrsim 4$~mag, the impact of destruction by dominant ions (e.g. \ce{H+}, \ce{H3+}, \ce{H3O+}, \ce{C+}, \ce{HCO+}) gradually increases with the optical depth, since UV-photon penetration diminishes and most of the ion abundances increase. 
Other destruction pathways involving atomic carbon forming bigger carbon chain molecules appear with increasing optical depth, but these are typically far less common in our grid models and are only relevant in dense clouds with higher carbon abundances. 

The formation of \chtcn is dominated by the dissociative recombination of CH$_3$CNH$^+$ with electrons, where CH$_3$CNH$^+$ itself is primarily formed by the radiative association~(\ref{eq:2}) and the following:
\begin{equation}
    \ce{HNC + CH3+ ->[$k_{13}$] CH3CNH+ + $h\nu$,}
\end{equation}
with $k_{13}$ given in Table~\ref{tab:new_reactions}.
Similarly, and as already presented in \S~\ref{subsec:ch3cn_ch3nc}, the formation of \ce{CH3NC} is dominated by the dissociative recombination of CH$_3$NCH$^+$ with electrons, with
CH$_3$NCH$^+$ primarily formed by the radiative association~(\ref{eq:1}).
\ce{CH3+} is formed via successive hydrogenation from \ce{C+}. HCN and HNC also descend from atomic carbon \citep{legal2014,loison2014}.
As a consequence, the formations of \chtcn and its isomer \ce{CH3NC} seem to be guided mostly by the carbon abundance, explaining that the abundance of \chtcn increases with C/O (see Fig.~\ref{fig:meudon_co}). 
The destruction of the isomers are dominated by photo-dissociation:
\begin{equation}
\ce{CH3CN/CH3NC + $h\nu$ ->[$k_{14}$] CH3 + CN},
\end{equation}
with $k_{14}$ given in Table~\ref{tab:new_reactions}.

To summarize, the enhancement in nitrile abundances appear to be tightly correlated with the C/O ratio. More interestingly however, our study highlights the importance of the relative elemental gas-phase abundances with respect to one another, and emphasizes the indirect role of oxygen in nitrile chemistry. The latter has a dramatic impact on the carbon chemistry in O-rich molecular environments, where most of the carbon is rapidly locked in CO and atomic O is a main destroyer for hydrocarbons, hampering the development of more complex carbon chemistry. In addition, we have shown in \S~\ref{subsec:C_O_impact} that the nitrile abundances increase with $\zeta$ (Fig.~\ref{fig:meudon_co}). Indeed, the ionization fraction is directly linked to $\zeta$, which governs the production of ions, such as \ce{CH3+}, \ce{HC3NH+} and \ce{c-C3H2+}, parent molecules of the complex nitriles here under study. \ce{CH3+} react with HCN and HNC to produce \ce{CH3CNH+}, that readily recombines with electrons to form \chtcn; \ce{HC3NH+} recombines with electrons to form \hctn (\ref{reac:hc3n_form1}); and \ce{c-C3H2+} recombines with electrons to form \ce{c-C3H} which reacts with N to produce \ctn~(\ref{reac:c3n_form}).

\subsection{The role of C/O in PDR and disk atmosphere nitrile chemistry}
\label{subsec:c_o_ratio_discussion}

The strong impact of the C/O ratio on the nitrile chemistry can likely be explained by the fact that in the gas phase, for a fixed amount of C element, O-removal decreases one of the main destruction pathways of hydrocarbons, which are the parent molecules of nitriles. To this effect adds the reactions of small carbon and oxygen species to form CO. Typically, in molecular clouds, most of the carbon hastily reacts with all available oxygen to form CO, effectively removing it from the reaction pathways that build up more complex molecules, such as complex nitriles. O-removal thus leaves more `free' carbon available in the gas phase to form carbon-containing molecules such as hydrocarbons, carbon chains and nitriles. As a comparison, the {\it standard} model results in a CO abundance of almost $1.38\times10^{-4}$, i.e. quasi all the carbon available in our models (see Table~\ref{tab:abun}), whereas our best-fit model produces a CO abundance of $\sim 1.34\times10^{-4}$. This leaves $\sim 6.0\times10^{-6}$ `free' carbon to build more complex carbon-containing molecules.

Varying the C/O ratio also impacts the gas temperature for \av$\lesssim1.5$~mag, as shown in the last panel row of Fig.~\ref{fig:meudon_co}. An increase in C/O ratio increases the gas temperature. This is due to the fact that, in the present work, we varied the abundance of atomic O to change the C/O ratio, thus reducing one of the main gas coolant in this region of the PDR. To test the impact of varying the C/O ratio via the carbon elemental abundance instead of the oxygen, we ran additional models. The main difference is in the resulting CO abundance. For a fixed C/O ratio, it increases with the amount of carbon. As a consequence, the gas temperature decreases at the edge of the PDR, since CO is another important gas coolant. However, these differences diminish with the increase of the visual extinction and are minor at the PDR position and in particular on the nitrile abundances. Thus, the main results found in the present work on the C/O impact on the Horsehead nebula chemistry is agnostic to whether C or O is varied to achieve a C/O ratio of $\sim$1. 

In disks, the C/O ratio is also strongly impacting the nitrile chemistry with the additional effect of changing the morphology of the nitrile abundance throughout the disk, as shown in Fig.~\ref{fig:disk_model_c_o_comp}. Whether or not \chtcn forms in gas or through gas-grain chemistry in disks, an elevated C/O in disks also helps in better reproducing the observations. However, grain chemistry seems to be needed to better reproduce the observations, in agreement with \cite{oberg2015nat} and \cite{loomis2018}.
Our finding of a C/O$\gtrsim1$ needed to reproduce the complex nitriles chemistry observed in disk atmospheres is in good agreement with the results of \cite{bergin2016} that also find that a C/O ratio exceeding unity is required to reproduce the observations of \ce{C2H} in disks. These authors therefore proposed \ce{C2H} as probe of C/O-enriched disk layers. Here we propose that complex nitriles could also serve as such probe in both PDR and disks, with the vantage of also probing the internal parts of the latter ($\lesssim 100$~au) as predicted by our models (see Fig.~\ref{fig:disk_model_c_o_comp}). Moreover, the fact that complex nitriles, such as \hctn and \chtcn, are much more commonly observed in disks than O-containing complex molecules, such as \ce{CH3OH}, highly suggests that organic chemistry is regulated by the C/O ratio and that disk atmosphere chemistry seems to be more generally C-rich than O-rich.

\section{Conclusions}
\label{sec:conclusion}

We tackled the chemistry of nitriles in PDR, in order to, first, understand the observations found for the nitriles \ctn, \hctn, \chtcn and \ce{CH3NC} toward the Horsehead PDR \citep{gratier2013}, and second, test if our improved PDR chemistry could help to explain the recent observations of \hctn and \chtcn in disk atmospheres \citep{oberg2015nat,bergner2018}. To this aim, 
we extended the chemistry of the \texttt{Meudon PDR code (v.1.5.2)} up to these four nitriles and explored the impact of some key parameters, such as the cosmic-ray ionization rate $\zeta$ and the gas-phase elemental C/O ratio, on our modeled nitrile results. Our main conclusions are summarized below:
\begin{enumerate}
    \item Varying the C/O ratio between 0.4 and 1.5 in a model of the Horsehead PDR results in orders of magnitude changes in nitrile abundances.
    \item A gas-phase C/O ratio of $\gtrsim0.9$ can reproduce the \ctn, \hctn, \chtcn and \ce{CH3NC} abundances within an order of magnitude in the Horsehead PDR, without any grain-surface chemistry.
    \item The cosmic-ray ionization rate moderately affects the nitrile chemistry through its impact on electrons and hydrocarbon ions (e.g. \ce{CH3+, \ce{C3H2+}}). A $\zeta=2\times10^{-16}\s$ per \ce{H2} better fits the observations than the standard value of $\zeta=5\times10^{-17}\s$ per \ce{H2}.
    \item Our best fit PDR model (i.e. with $\zeta=2\times10^{-16}\s$ per \ce{H2} and C/O$=1$) can reproduce both the relative abundances of \ce{CH3NC} and \ce{CH3CN} when adopting a branching ratio of 0.8 isomerization for the reaction \ce{CH3+ + HCN} (\ref{eq:2}).
    \item An elevated C/O ratio ($\sim1.0$) could also be the key for understanding complex nitrile disk chemistry. Using a fiducial protoplanetary disk astrochemical model, we find that disk observations of \chtcn and \hctn are reproduced within an order of magnitude, while our standard model under-predicts the same molecules by 2-3 orders of magnitude.\\
\end{enumerate}

While the good agreement between observations and models in both a classic PDR and a planet-forming disk, it is important to note that the nitrile grain chemistry is still poorly constrained, and it may contribute to both kinds of regions \citep[e.g.][]{bertin2017a,bertin2017b,calcutt2018,nguyen2019}. Further experiments and theory on nitrile grain surface chemistry and desorption are needed to make progress here. Thus, it would be interesting to add grain chemistry, and in particular grain nitrile chemistry, in the \texttt{Meudon PDR code} to test how the results presented here would be affected.
Furthermore, other parameters would be worth testing in future model developments, such as the impact of stellar X-ray irradiation on disk chemistry which could affect its ionization \citep[e.g.][]{glassgold1997,rab2018,waggoner2019}, and the sensitivity of disk chemistry to grain sizes \citep[e.g.][]{wakelam2019} as smaller grains provide a higher surface area relative to their volume and thus more reaction sites and also tend to have temperatures closer to that of the gas.
In the meantime we note that the strong impact of C/O on nitrile chemistry may enable us to use nitriles to constrain this important parameter in disks and PDR analogs. 

\vspace{-6mm}
\acknowledgments 
The authors would like to thank the anonymous referee for constructive suggestions and comments.
R.~LG. also thanks Tom J. Millar for useful discussion. This work was supported by an award from the Simons Foundation (SCOL \# 321183, KO). M.~T.~B.\ acknowledges support from the Noland Internship at Caltech.\\\\

\vspace{-8mm}
\software 
{\texttt{Pandas} \citep{pandas}, \texttt{Matplotlib} \citep{matplotlib}, \texttt{NumPy} \citep{numpy_article}, \texttt {SciPy} \citep{scipy}, \texttt{Meudon PDR Code} \citep{lebourlot1993,lepetit2006,lebourlot2012}, \texttt{Nautilus-v1.1} \citep{hersant2009,ruaud2016,wakelam2016}.}
\newline

\appendix
\vspace{-3mm}
\restartappendixnumbering
\section{Critical chemical reactions added to the Meudon PDR Code network}
\begin{table}[!h]
    \scriptsize
    \centering
    \caption{Rates of the critical chemical reactions discussed in this study. \label{tab:new_reactions}}
    \begin{tabular}{ccccccccccccc}
    \hline
    \hline
    \multicolumn{6}{c}{Chemical reactions}& $\alpha$&$\beta$ & $\gamma$& $k$ rate  & $T$ range& Ref.$^{(b)}$ & Rate \\
       &&&&&&($\rm{cm}^{3}\s$)&&&type$^{(a)}$&(K)&&uncertainties$^{(c)}$\\
    \hline
    \multicolumn{12}{c}{Main formation and destruction pathways for \ce{CH3CNH+} \& \ce{CH3CN}}\\
     \cline{4-10}
    \ce{CH3+}&\ce{HCN}&$\rightarrow$&\ce{CH3CNH+}&$h\nu$  & &  7.20(-9) & -0.50 & 0.00 &  (1) & $10-299$&[1], [2]& $F_0=10,\, g=0$ \\
    \ce{CH3+}&\ce{HCN}&$\rightarrow$&\ce{CH3CNH+}&$h\nu$  & &  8.00(-11) &-3.00 & 0.00 &  (1) & $300-800$& [2], [3]&$F_0=2,\, g=0$ \\
    \ce{CH3+}&\ce{HNC}&$\rightarrow$&\ce{CH3CNH+}&$h\nu$  & &  9.00(-9) & -0.50 & 0.00 &  (1) & $10-300$& [4]&$F_0=10,\, g=0$ \\
    \ce{CH3CNH+}&\ce{e-}&$\rightarrow$&\ce{H}&\ce{H}&\ce{H2CCN}&  8.00(-8) &-0.50 &    0.00 &  (2) & $10-300$& [4]&$F_0=3,\, g=0$\\
    \ce{CH3CNH+}&\ce{e-}&$\rightarrow$&\ce{H}&\ce{CH3CN}& &  1.30(-7) &-0.50 &    0.00 &  (2) & $10-300$&[4]&$F_0=3,\, g=0$\\
    \ce{CH3CNH+}&\ce{e-}&$\rightarrow$&\ce{HNC}&\ce{CH3}& &  6.00(-8) &-0.50 &    0.00 &  (2) & $10-300$&[4]&$F_0=3,\, g=0$\\
    \ce{CH3CNH+}&\ce{e-}&$\rightarrow$&\ce{HCN}&\ce{CH3}& &  6.00(-8) &-0.50 &    0.00 &  (2) & $10-300$&[4]&$F_0=3,\, g=0$\\
    \ce{CH3CN}&$h\nu$ &$\rightarrow$&\ce{CN} & \ce{CH3}    & &  2.95(-9) & 0.00 &    3.07 &  (3) &-- &[5]&$F_0=1.3,\, g=0$\\
    \hline
    \multicolumn{12}{c}{Main formation and destruction pathways for \ce{CH3NCH+} \& \ce{CH3NC}}\\
    \cline{4-10}
    \ce{CH3+}&\ce{HCN}&$\rightarrow$&\ce{CH3NCH+}&$h\nu$  & &  1.80(-9) & -0.50 & 0.00 &  (1) & $10-299$&[1], [2]&$F_0=10,\, g=0$\\
    \ce{CH3+}&\ce{HCN}&$\rightarrow$&\ce{CH3NCH+}&$h\nu$  & &  2.00(-11) &-3.00 & 0.00 &  (1) & $300-800$&[2], [3]&$F_0=2,\, g=0$\\
    \ce{CH3NCH+}&\ce{e-}&$\rightarrow$&\ce{H}&\ce{CH3NC}& &  1.30(-7) &-0.50 &    0.00 &  (2) & $10-300$& [4]$^{(\star)}$&$F_0=3,\, g=0$\\
    \ce{CH3NCH+}&\ce{e-}&$\rightarrow$&\ce{HCN}&\ce{CH3}& &  6.00(-8) &-0.50 &    0.00 &  (2) & $10-300$& [4]$^{(\star)}$&$F_0=3,\, g=0$\\
    \ce{CH3NC}&$h\nu$ &$\rightarrow$&\ce{CN}&\ce{CH3}& &  2.95(-9) & 0.00 &    3.07 &  (3) & -- & [5]$^{(\star)}$ &$F_0=1.3,\, g=0$\\
    \hline
    \multicolumn{12}{c}{Main formation and destruction pathways for \ce{C3N}}\\
    \cline{4-10}
    \ce{N}&\ce{c-C3H}&$\rightarrow$&\ce{H}&\ce{C3N}& &  1.10(-10) & 0.17 &    0.00 &  (2) &$10-300$ & [4]&$F_0=3,\, g=0$\\
    \ce{C3N}&$h\nu$&$\rightarrow$&\ce{C2}&\ce{CN}& &  5.00(-10) & 0.00 &    1.80 &  (3) & -- &[6]&$F_0=2,\, g=0$\\
    \ce{C3N}&\ce{O}&$\rightarrow$&\ce{CO}&\ce{CCN}& &  1.00(-10) & 0.00 &    0.00 &  (2) & $10-298$ & [7]&$F_0=3,\, g=2.97$\\
    \ce{C3N} &\ce{C}&$\rightarrow$&\ce{CN}&\ce{C3}& &  2.40(-10) & 0.00 &    0.00 &  (2) & $10-300$&[4]&$F_0=3,\, g=0$\\
    \hline
    \multicolumn{12}{c}{Main formation and destruction pathways for \ce{HC3N}}\\
    \cline{4-10}
    \ce{HC3NH+}&\ce{e-}&$\rightarrow$&\ce{H}&\ce{HC3N}& &  6.00(-7) &-0.58 &    0.00 &  (2) & $10-800$&[8]&$F_0=1.4,\, g=0$\\
    \ce{C4H}&\ce{N}&$\rightarrow$&\ce{C}&\ce{HC3N}& &  7.00(-11) & 0.17 &    0.00 &  (2) & $10-300$ &[4]&$F_0=3,\, g=0$\\
    \ce{C}&\ce{H2CCN}&$\rightarrow$&\ce{H}&\ce{HC3N}& &  1.00(-10) & 0.00 &    0.00 &  (2) &$10-300$ &[4]&$F_0=3,\, g=0$\\
    \ce{N}&\ce{c-C3H2}&$\rightarrow$&\ce{H}&\ce{HC3N}& &  1.00(-11) & 0.00 & 2000.00 &  (2) & $10-800$ &[8]&$F_0=3,\, g=500$\\
    \ce{C2H2}&\ce{CN}&$\rightarrow$&\ce{H}&\ce{HC3N}& &  2.72(-10) &-0.52 &   19.00 &  (2) & $10-280$& [6]&$F_0=1.25,\, g=0$\\
    \ce{HC3N}&$h\nu$&$\rightarrow$&\ce{CN}&\ce{C2H}& &  7.13(-9) & 0.00 & 2.59 & (3) & -- &[5]&$F_0=1.3,\, g=0$\\
    \hline
    \hline
    \end{tabular}
    \tablecomments{Numbers in parentheses are power of 10 ${(a)}$ Rate formulae: (1) Radiative associations and (2) bimolecular gas-phase reactions are computed from the same rate formula $k = \alpha (T/300)^{\beta} e^{-\gamma/T}$, (3) photo-reactions for which the photo-cross-sections are unknown are computed with
         $k = \alpha \left( \chi^{-} e^{-\gamma A_{\rm{V}}}  + \chi^{+} e^{-\gamma    
        (A_{\rm{V}}^{\rm{max}} - A_{\rm{V}})}\right)$, 
         with $\chi^{-}$ and $\chi^{+}$ scaling factors of the radiation field with respect to that of Draine on the left and right side of the cloud, respectively \citep{lepetit2006}; ${(b)}$ [1] \cite{herbst1985}; [2] \cite{anicich2003}; [3] \cite{harada2010}; [4] \cite{loison2014}; [5] \cite{heays2017}; [6] OSU database; [7] Smith database; [8] \cite{loison2017}; $(\star)$ Assumed from the referenced publication; ${(c)}$ Rate uncertainties are from the KIDA database and are considered to follow a lognormal distribution, i.e. with a probability of $\approx 68$\% for the rate $k$ to be in the range $[\frac{k}{F_0}, k \times F_0]$ and $g$ being an expansion parameter used to parameterize a possible temperature dependence of the uncertainty, according to the formula $F(T)=F_0\exp(g|\frac{1}{T}-\frac{1}{T_0}|)$ with $T_0=300$~K \citep[see][for more details]{wakelam2012}.}  
\end{table}

\newpage
\bibliography{manuscript}

\begin{thebibliography}{}
\expandafter\ifx\csname natexlab\endcsname\relax\def\natexlab#1{#1}\fi
\providecommand{\url}[1]{\href{#1}{#1}}

\bibitem[{{Abergel} {et~al.}(2003){Abergel}, {Teyssier}, {Bernard},
  {Boulanger}, {Coulais}, {Fosse}, {Falgarone}, {Gerin}, {Perault}, {Puget},
  {Nordh}, {Olofsson}, {Huldtgren}, {Kaas}, {Andr{\'e}}, {Bontemps}, {Casali},
  {Cesarsky}, {Copet}, {Davies}, {Montmerle}, {Persi}, \&
  {Sibille}}]{abergel2003}
{Abergel}, A., {Teyssier}, D., {Bernard}, J.~P., {et~al.} 2003, \aap, 410, 577

\bibitem[{{Andrews} {et~al.}(2013){Andrews}, {Rosenfeld}, {Kraus}, \&
  {Wilner}}]{andrews2013}
{Andrews}, S.~M., {Rosenfeld}, K.~A., {Kraus}, A.~L., \& {Wilner}, D.~J. 2013,
  \apj, 771, 129

\bibitem[{Anicich(2003)}]{anicich2003}
Anicich, V.~G. 2003, JPL Publication 03-19, 1

\bibitem[{{Anicich} {et~al.}(1995){Anicich}, {Sen}, {Huntress}, \&
  {McEwan}}]{anicich1995}
{Anicich}, V.~G., {Sen}, A.~D., {Huntress}, Wesley~T., J., \& {McEwan}, M.~J.
  1995, \jcp, 102, 3256

\bibitem[{{Asplund} {et~al.}(2009){Asplund}, {Grevesse}, {Sauval}, \&
  {Scott}}]{asplund2009}
{Asplund}, M., {Grevesse}, N., {Sauval}, A.~J., \& {Scott}, P. 2009, \araa, 47,
  481

\bibitem[{{Beckwith} {et~al.}(1990){Beckwith}, {Sargent}, {Chini}, \&
  {Guesten}}]{beckwith1990}
{Beckwith}, S.~V.~W., {Sargent}, A.~I., {Chini}, R.~S., \& {Guesten}, R. 1990,
  \aj, 99, 924

\bibitem[{{Bergin} {et~al.}(2016){Bergin}, {Du}, {Cleeves}, {Blake}, {Schwarz},
  {Visser}, \& {Zhang}}]{bergin2016}
{Bergin}, E.~A., {Du}, F., {Cleeves}, L.~I., {et~al.} 2016, \apj, 831, 101

\bibitem[{{Bergner} {et~al.}(2018){Bergner}, {Guzm{\'a}n}, {{\"O}berg},
  {Loomis}, \& {Pegues}}]{bergner2018}
{Bergner}, J.~B., {Guzm{\'a}n}, V.~G., {{\"O}berg}, K.~I., {Loomis}, R.~A., \&
  {Pegues}, J. 2018, \apj, 857, 69

\bibitem[{{Bergner} {et~al.}(2019){Bergner}, {{\"O}berg}, {Bergin}, {Loomis},
  {Pegues}, \& {Qi}}]{bergner2019}
{Bergner}, J.~B., {{\"O}berg}, K.~I., {Bergin}, E.~A., {et~al.} 2019, \apj,
  876, 25

\bibitem[{{Bertin} {et~al.}(2017{\natexlab{a}}){Bertin}, {Doronin}, {Fillion},
  {Michaut}, {Philippe}, {Lattelais}, {Markovits}, {Pauzat}, {Ellinger}, \&
  {Guillemin}}]{bertin2017a}
{Bertin}, M., {Doronin}, M., {Fillion}, J.~H., {et~al.} 2017{\natexlab{a}},
  \aap, 598, A18

\bibitem[{{Bertin} {et~al.}(2017{\natexlab{b}}){Bertin}, {Doronin}, {Michaut},
  {Philippe}, {Markovits}, {Fillion}, {Pauzat}, {Ellinger}, \&
  {Guillemin}}]{bertin2017b}
{Bertin}, M., {Doronin}, M., {Michaut}, X., {et~al.} 2017{\natexlab{b}}, \aap,
  608, A50

\bibitem[{{Calcutt} {et~al.}(2018){Calcutt}, {Fiechter}, {Willis},
  {M{\"u}ller}, {Garrod}, {J{\o}rgensen}, {Wampfler}, {Bourke}, {Coutens}, \&
  {Drozdovskaya}}]{calcutt2018}
{Calcutt}, H., {Fiechter}, M.~R., {Willis}, E.~R., {et~al.} 2018, \aap, 617,
  A95

\bibitem[{{Cernicharo} {et~al.}(1988){Cernicharo}, {Kahane}, {Guelin}, \&
  {Gomez-Gonzalez}}]{cernicharo1988}
{Cernicharo}, J., {Kahane}, C., {Guelin}, M., \& {Gomez-Gonzalez}, J. 1988,
  \aap, 189, L1

\bibitem[{{Chapillon} {et~al.}(2012){Chapillon}, {Dutrey}, {Guilloteau},
  {Pi{\'e}tu}, {Wakelam}, {Hersant}, {Gueth}, {Henning}, {Launhardt},
  {Schreyer}, \& {Semenov}}]{chapillon2012}
{Chapillon}, E., {Dutrey}, A., {Guilloteau}, S., {et~al.} 2012, \apj, 756, 58

\bibitem[{{Chiang} \& {Goldreich}(1997)}]{chiang1997}
{Chiang}, E.~I., \& {Goldreich}, P. 1997, \apj, 490, 368

\bibitem[{{Cleeves} {et~al.}(2015){Cleeves}, {Bergin}, {Qi}, {Adams}, \&
  {{\"O}berg}}]{cleeves2015}
{Cleeves}, L.~I., {Bergin}, E.~A., {Qi}, C., {Adams}, F.~C., \& {{\"O}berg},
  K.~I. 2015, \apj, 799, 204

\bibitem[{{Cleeves} {et~al.}(2018){Cleeves}, {{\"O}berg}, {Wilner}, {Huang},
  {Loomis}, {Andrews}, \& {Guzman}}]{cleeves2018}
{Cleeves}, L.~I., {{\"O}berg}, K.~I., {Wilner}, D.~J., {et~al.} 2018, \apj,
  865, 155

\bibitem[{{Cuadrado} {et~al.}(2015){Cuadrado}, {Goicoechea}, {Pilleri},
  {Cernicharo}, {Fuente}, \& {Joblin}}]{cuadrado2015}
{Cuadrado}, S., {Goicoechea}, J.~R., {Pilleri}, P., {et~al.} 2015, \aap, 575,
  A82

\bibitem[{{Dartois} {et~al.}(2003){Dartois}, {Dutrey}, \&
  {Guilloteau}}]{dartois2003}
{Dartois}, E., {Dutrey}, A., \& {Guilloteau}, S. 2003, \aap, 399, 773

\bibitem[{{Defrees} {et~al.}(1985){Defrees}, {McLean}, \&
  {Herbst}}]{defrees1985}
{Defrees}, D.~J., {McLean}, A.~D., \& {Herbst}, E. 1985, \apj, 293, 236

\bibitem[{{Draine}(1978)}]{draine1978}
{Draine}, B.~T. 1978, \apjs, 36, 595

\bibitem[{{Dullemond} {et~al.}(2001){Dullemond}, {Dominik}, \&
  {Natta}}]{dullemond2001}
{Dullemond}, C.~P., {Dominik}, C., \& {Natta}, A. 2001, \apj, 560, 957

\bibitem[{{Dutrey} {et~al.}(1997){Dutrey}, {Guilloteau}, \&
  {Guelin}}]{dutrey1997}
{Dutrey}, A., {Guilloteau}, S., \& {Guelin}, M. 1997, \aap, 317, L55

\bibitem[{{Elmegreen}(2000)}]{emelgreen2000}
{Elmegreen}, B.~G. 2000, \apj, 530, 277

\bibitem[{{Fuente} {et~al.}(2010){Fuente}, {Cernicharo}, {Ag{\'u}ndez},
  {Bern{\'e}}, {Goicoechea}, {Alonso-Albi}, \& {Marcelino}}]{fuente2010}
{Fuente}, A., {Cernicharo}, J., {Ag{\'u}ndez}, M., {et~al.} 2010, \aap, 524,
  A19

\bibitem[{{Fuente} {et~al.}(2003){Fuente}, {Rodr{\i}guez-Franco},
  {Garc{\i}a-Burillo}, {Mart{\i}n-Pintado}, \& {Black}}]{fuente2003}
{Fuente}, A., {Rodr{\i}guez-Franco}, A., {Garc{\i}a-Burillo}, S.,
  {Mart{\i}n-Pintado}, J., \& {Black}, J.~H. 2003, \aap, 406, 899

\bibitem[{{Gerin} {et~al.}(2009){Gerin}, {Goicoechea}, {Pety}, \&
  {Hily-Blant}}]{gerin2009}
{Gerin}, M., {Goicoechea}, J.~R., {Pety}, J., \& {Hily-Blant}, P. 2009, \aap,
  494, 977

\bibitem[{{Glassgold} \& {Langer}(1974)}]{glassgold1974}
{Glassgold}, A.~E., \& {Langer}, W.~D. 1974, \apj, 193, 73

\bibitem[{{Glassgold} {et~al.}(1997){Glassgold}, {Najita}, \&
  {Igea}}]{glassgold1997}
{Glassgold}, A.~E., {Najita}, J., \& {Igea}, J. 1997, \apj, 480, 344

\bibitem[{{Goicoechea} {et~al.}(2009){Goicoechea}, {Pety}, {Gerin},
  {Hily-Blant}, \& {Le Bourlot}}]{goicoechea2009}
{Goicoechea}, J.~R., {Pety}, J., {Gerin}, M., {Hily-Blant}, P., \& {Le
  Bourlot}, J. 2009, \aap, 498, 771

\bibitem[{{Goicoechea} {et~al.}(2006){Goicoechea}, {Pety}, {Gerin}, {Teyssier},
  {Roueff}, {Hily-Blant}, \& {Baek}}]{goicoechea2006}
{Goicoechea}, J.~R., {Pety}, J., {Gerin}, M., {et~al.} 2006, \aap, 456, 565

\bibitem[{{Gratier} {et~al.}(2013){Gratier}, {Pety}, {Guzm{\'a}n}, {Gerin},
  {Goicoechea}, {Roueff}, \& {Faure}}]{gratier2013}
{Gratier}, P., {Pety}, J., {Guzm{\'a}n}, V., {et~al.} 2013, \aap, 557, A101

\bibitem[{{Grenier} {et~al.}(2015){Grenier}, {Black}, \&
  {Strong}}]{grenier2015}
{Grenier}, I.~A., {Black}, J.~H., \& {Strong}, A.~W. 2015, \araa, 53, 199

\bibitem[{{Guilloteau} {et~al.}(2011){Guilloteau}, {Dutrey}, {Pi{\'e}tu}, \&
  {Boehler}}]{guilloteau2011}
{Guilloteau}, S., {Dutrey}, A., {Pi{\'e}tu}, V., \& {Boehler}, Y. 2011, \aap,
  529, A105

\bibitem[{{Guilloteau} {et~al.}(2016){Guilloteau}, {Reboussin}, {Dutrey},
  {Chapillon}, {Wakelam}, {Pi{\'e}tu}, {Di Folco}, {Semenov}, \&
  {Henning}}]{guilloteau2016}
{Guilloteau}, S., {Reboussin}, L., {Dutrey}, A., {et~al.} 2016, \aap, 592, A124

\bibitem[{{Guzm{\'a}n} {et~al.}(2011){Guzm{\'a}n}, {Pety}, {Goicoechea},
  {Gerin}, \& {Roueff}}]{guzman2011}
{Guzm{\'a}n}, V., {Pety}, J., {Goicoechea}, J.~R., {Gerin}, M., \& {Roueff}, E.
  2011, \aap, 534, A49

\bibitem[{{Guzm{\'a}n} {et~al.}(2015){Guzm{\'a}n}, {Pety}, {Goicoechea},
  {Gerin}, {Roueff}, {Gratier}, \& {{\"O}berg}}]{guzman2015}
{Guzm{\'a}n}, V.~V., {Pety}, J., {Goicoechea}, J.~R., {et~al.} 2015, \apjl,
  800, L33

\bibitem[{{Habart} {et~al.}(2005){Habart}, {Abergel}, {Walmsley}, {Teyssier},
  \& {Pety}}]{habart2005}
{Habart}, E., {Abergel}, A., {Walmsley}, C.~M., {Teyssier}, D., \& {Pety}, J.
  2005, \aap, 437, 177

\bibitem[{Harada {et~al.}(2010)Harada, Herbst, \& Wakelam}]{harada2010}
Harada, N., Herbst, E., \& Wakelam, V. 2010, Astrophysical Journal, 721, 1570

\bibitem[{{Hartmann} {et~al.}(2001){Hartmann}, {Ballesteros-Paredes}, \&
  {Bergin}}]{hartmann2001}
{Hartmann}, L., {Ballesteros-Paredes}, J., \& {Bergin}, E.~A. 2001, \apj, 562,
  852

\bibitem[{{Heays} {et~al.}(2017){Heays}, {Bosman}, \& {van
  Dishoeck}}]{heays2017}
{Heays}, A.~N., {Bosman}, A.~D., \& {van Dishoeck}, E.~F. 2017, \aap, 602, A105

\bibitem[{{Henning} {et~al.}(2010){Henning}, {Semenov}, {Guilloteau}, {Dutrey},
  {Hersant}, {Wakelam}, {Chapillon}, {Launhardt}, {Pi{\'e}tu}, \&
  {Schreyer}}]{henning2010}
{Henning}, T., {Semenov}, D., {Guilloteau}, S., {et~al.} 2010, \apj, 714, 1511

\bibitem[{Herbst(1985)}]{herbst1985}
Herbst, E. 1985, Astrophysical Journal, 291, 226

\bibitem[{{Hersant} {et~al.}(2009){Hersant}, {Wakelam}, {Dutrey}, {Guilloteau},
  \& {Herbst}}]{hersant2009}
{Hersant}, F., {Wakelam}, V., {Dutrey}, A., {Guilloteau}, S., \& {Herbst}, E.
  2009, \aap, 493, L49

\bibitem[{{Hogerheijde} {et~al.}(2011){Hogerheijde}, {Bergin}, {Brinch},
  {Cleeves}, {Fogel}, {Blake}, {Dominik}, {Lis}, {Melnick}, {Neufeld},
  {Pani{\'c}}, {Pearson}, {Kristensen}, {Y{\i}ld{\i}z}, \& {van
  Dishoeck}}]{hogerheijde2011}
{Hogerheijde}, M.~R., {Bergin}, E.~A., {Brinch}, C., {et~al.} 2011, Science,
  334, 338

\bibitem[{{Huang} {et~al.}(2018){Huang}, {Andrews}, {Dullemond}, {Isella},
  {P{\'e}rez}, {Guzm{\'a}n}, {{\"O}berg}, {Zhu}, {Zhang}, {Bai}, {Benisty},
  {Birnstiel}, {Carpenter}, {Hughes}, {Ricci}, {Weaver}, \&
  {Wilner}}]{huang2018}
{Huang}, J., {Andrews}, S.~M., {Dullemond}, C.~P., {et~al.} 2018, arXiv
  e-prints, arXiv:1812.04041

\bibitem[{Hunter(2007)}]{matplotlib}
Hunter, J.~D. 2007, Computing in Science Engineering, 9, 90

\bibitem[{{Huntress} \& {Mitchell}(1979)}]{huntress1979}
{Huntress}, W.~T., J., \& {Mitchell}, G.~F. 1979, \apj, 231, 456

\bibitem[{{Indriolo} {et~al.}(2015){Indriolo}, {Neufeld}, {Gerin}, {Schilke},
  {Benz}, {Winkel}, {Menten}, {Chambers}, {Black}, {Bruderer}, {Falgarone},
  {Godard}, {Goicoechea}, {Gupta}, {Lis}, {Ossenkopf}, {Persson},
  {Sonnentrucker}, {van der Tak}, {van Dishoeck}, {Wolfire}, \&
  {Wyrowski}}]{indriolo2015}
{Indriolo}, N., {Neufeld}, D.~A., {Gerin}, M., {et~al.} 2015, \apj, 800, 40

\bibitem[{{Jenkins}(2009)}]{jenkins2009}
{Jenkins}, E.~B. 2009, \apj, 700, 1299

\bibitem[{{Jones} \& {Ysrad}(2019)}]{jones2019}
{Jones}, A.~P., \& {Ysrad}, N. 2019, arXiv e-prints, arXiv:1906.01382

\bibitem[{Jones {et~al.}(2001--)Jones, Oliphant, Peterson, {et~al.}}]{scipy}
Jones, E., Oliphant, T., Peterson, P., {et~al.} 2001--, SciPy: Open source
  scientific tools for Python, , .
\newblock \url{http://www.scipy.org/}

\bibitem[{{Kastner} {et~al.}(2015){Kastner}, {Qi}, {Gorti}, {Hily-Blant},
  {Oberg}, {Forveille}, {Andrews}, \& {Wilner}}]{kastner2015}
{Kastner}, J.~H., {Qi}, C., {Gorti}, U., {et~al.} 2015, \apj, 806, 75

\bibitem[{{Kastner} {et~al.}(2018){Kastner}, {Qi}, {Dickson-Vandervelde},
  {Hily-Blant}, {Forveille}, {Andrews}, {Gorti}, {{\"O}berg}, \&
  {Wilner}}]{kastner2018}
{Kastner}, J.~H., {Qi}, C., {Dickson-Vandervelde}, D.~A., {et~al.} 2018, \apj,
  863, 106

\bibitem[{{Le Bourlot} {et~al.}(2012){Le Bourlot}, {Le Petit}, {Pinto},
  {Roueff}, \& {Roy}}]{lebourlot2012}
{Le Bourlot}, J., {Le Petit}, F., {Pinto}, C., {Roueff}, E., \& {Roy}, F. 2012,
  \aap, 541, A76

\bibitem[{{Le Bourlot} {et~al.}(1993){Le Bourlot}, {Pineau Des Forets},
  {Roueff}, \& {Flower}}]{lebourlot1993}
{Le Bourlot}, J., {Pineau Des Forets}, G., {Roueff}, E., \& {Flower}, D.~R.
  1993, \aap, 267, 233

\bibitem[{{Le Gal} {et~al.}(2017){Le Gal}, {Herbst}, {Dufour}, {Gratier},
  {Ruaud}, {Vidal}, \& {Wakelam}}]{legal2017}
{Le Gal}, R., {Herbst}, E., {Dufour}, G., {et~al.} 2017, \aap, 605, A88

\bibitem[{{Le Gal} {et~al.}(2014){Le Gal}, {Hily-Blant}, {Faure}, {Pineau des
  For{\^e}ts}, {Rist}, \& {Maret}}]{legal2014}
{Le Gal}, R., {Hily-Blant}, P., {Faure}, A., {et~al.} 2014, \aap, 562, A83

\bibitem[{{Le Gal} {et~al.}(2019){Le Gal}, {{\"O}berg}, {Loomis}, {Pegues}, \&
  {Bergner}}]{legal2019}
{Le Gal}, R., {{\"O}berg}, K.~I., {Loomis}, R.~A., {Pegues}, J., \& {Bergner},
  J.~B. 2019, \apj, 876, 72

\bibitem[{{Le Petit} {et~al.}(2006){Le Petit}, {Nehm{\'e}}, {Le Bourlot}, \&
  {Roueff}}]{lepetit2006}
{Le Petit}, F., {Nehm{\'e}}, C., {Le Bourlot}, J., \& {Roueff}, E. 2006, \apjs,
  164, 506

\bibitem[{{Le Petit} {et~al.}(2016){Le Petit}, {Ruaud}, {Bron}, {Godard},
  {Roueff}, {Languignon}, \& {Le Bourlot}}]{lepetit2016}
{Le Petit}, F., {Ruaud}, M., {Bron}, E., {et~al.} 2016, \aap, 585, A105

\bibitem[{{Loison} {et~al.}(2014){Loison}, {Wakelam}, \&
  {Hickson}}]{loison2014}
{Loison}, J.-C., {Wakelam}, V., \& {Hickson}, K.~M. 2014, \mnras, 443, 398

\bibitem[{{Loison} {et~al.}(2017){Loison}, {Ag{\'u}ndez}, {Wakelam}, {Roueff},
  {Gratier}, {Marcelino}, {Reyes}, {Cernicharo}, \& {Gerin}}]{loison2017}
{Loison}, J.-C., {Ag{\'u}ndez}, M., {Wakelam}, V., {et~al.} 2017, \mnras, 470,
  4075

\bibitem[{{Loomis} {et~al.}(2019){Loomis}, {Oberg}, {Andrews}, \& {et
  al.}}]{loomis2019}
{Loomis}, R., {Oberg}, K.~I., {Andrews}, S., \& {et al.} 2019, submitted to
  \apj

\bibitem[{{Loomis} {et~al.}(2018){Loomis}, {Cleeves}, {{\"O}berg}, {Aikawa},
  {Bergner}, {Furuya}, {Guzman}, \& {Walsh}}]{loomis2018}
{Loomis}, R.~A., {Cleeves}, L.~I., {{\"O}berg}, K.~I., {et~al.} 2018, \apj,
  859, 131

\bibitem[{{L{\'o}pez} {et~al.}(2014){L{\'o}pez}, {Tercero}, {Kisiel}, {Daly},
  {Berm{\'u}dez}, {Calcutt}, {Marcelino}, {Viti}, {Drouin}, \&
  {Medvedev}}]{lopez2014}
{L{\'o}pez}, A., {Tercero}, B., {Kisiel}, Z., {et~al.} 2014, \aap, 572, A44

\bibitem[{{Mathis} {et~al.}(1977){Mathis}, {Rumpl}, \&
  {Nordsieck}}]{mathis1977}
{Mathis}, J.~S., {Rumpl}, W., \& {Nordsieck}, K.~H. 1977, \apj, 217, 425

\bibitem[{McKinney(2010)}]{pandas}
McKinney, W. 2010, in Proceedings of the 9th Python in Science Conference, ed.
  S.~van~der Walt \& J.~Millman, 51 -- 56

\bibitem[{{Millar} \& {Herbst}(1990)}]{millar1990}
{Millar}, T.~J., \& {Herbst}, E. 1990, \mnras, 242, 92

\bibitem[{{Millar} {et~al.}(1987){Millar}, {Leung}, \& {Herbst}}]{millar1987}
{Millar}, T.~J., {Leung}, C.~M., \& {Herbst}, E. 1987, \aap, 183, 109

\bibitem[{{Nath} \& {Biermann}(1994)}]{nath1994}
{Nath}, B.~B., \& {Biermann}, P.~L. 1994, \mnras, 270, L33

\bibitem[{{Nguyen} {et~al.}(2019){Nguyen}, {Fourr{\'e}}, {Favre}, {Barois},
  {Congiu}, {Baouche}, {Guillemin}, {Ellinger}, \& {Dulieu}}]{nguyen2019}
{Nguyen}, T., {Fourr{\'e}}, I., {Favre}, C., {et~al.} 2019, \aap, 628, A15

\bibitem[{{{\"O}berg} {et~al.}(2015){{\"O}berg}, {Guzm{\'a}n}, {Furuya}, {Qi},
  {Aikawa}, {Andrews}, {Loomis}, \& {Wilner}}]{oberg2015nat}
{{\"O}berg}, K.~I., {Guzm{\'a}n}, V.~V., {Furuya}, K., {et~al.} 2015, \nat,
  520, 198

\bibitem[{Patel {et~al.}(2015)Patel, Percivalle, Ritson, Duffy, \&
  Sutherland}]{patel2015}
Patel, B.~H., Percivalle, C., Ritson, D.~J., Duffy, C.~D., \& Sutherland, J.~D.
  2015, Nature Chemistry, 7, 301 EP .
\newblock \url{https://doi.org/10.1038/nchem.2202}

\bibitem[{{Pety} {et~al.}(2007){Pety}, {Goicoechea}, {Hily-Blant}, {Gerin}, \&
  {Teyssier}}]{pety2007}
{Pety}, J., {Goicoechea}, J.~R., {Hily-Blant}, P., {Gerin}, M., \& {Teyssier},
  D. 2007, \aap, 464, L41

\bibitem[{{Pety} {et~al.}(2005){Pety}, {Teyssier}, {Foss{\'e}}, {Gerin},
  {Roueff}, {Abergel}, {Habart}, \& {Cernicharo}}]{pety2005}
{Pety}, J., {Teyssier}, D., {Foss{\'e}}, D., {et~al.} 2005, \aap, 435, 885

\bibitem[{{Pety} {et~al.}(2012){Pety}, {Gratier}, {Guzm{\'a}n}, {Roueff},
  {Gerin}, {Goicoechea}, {Bardeau}, {Sievers}, {Le Petit}, {Le Bourlot},
  {Belloche}, \& {Talbi}}]{pety2012}
{Pety}, J., {Gratier}, P., {Guzm{\'a}n}, V., {et~al.} 2012, \aap, 548, A68

\bibitem[{{Pi{\'e}tu} {et~al.}(2007){Pi{\'e}tu}, {Dutrey}, \&
  {Guilloteau}}]{pietu2007}
{Pi{\'e}tu}, V., {Dutrey}, A., \& {Guilloteau}, S. 2007, \aap, 467, 163

\bibitem[{{Plessis} {et~al.}(2012){Plessis}, {Carrasco}, {Dobrijevic}, \&
  {Pernot}}]{plessis2012}
{Plessis}, S., {Carrasco}, N., {Dobrijevic}, M., \& {Pernot}, P. 2012, \icarus,
  219, 254

\bibitem[{{Plessis} {et~al.}(2010){Plessis}, {Carrasco}, \&
  {Pernot}}]{plessis2010}
{Plessis}, S., {Carrasco}, N., \& {Pernot}, P. 2010, \jcp, 133, 134110

\bibitem[{Powner {et~al.}(2009)Powner, Gerland, \& Sutherland}]{Powner2009}
Powner, M.~W., Gerland, B., \& Sutherland, J.~D. 2009, Nature, 459, 239 EP .
\newblock \url{http://dx.doi.org/10.1038/nature08013}

\bibitem[{{Przybilla} {et~al.}(2008){Przybilla}, {Nieva}, \&
  {Butler}}]{przybilla2008}
{Przybilla}, N., {Nieva}, M.-F., \& {Butler}, K. 2008, \apjl, 688, L103

\bibitem[{{Qi} {et~al.}(2013){Qi}, {{\"O}berg}, {Wilner}, \&
  {Rosenfeld}}]{qi2013}
{Qi}, C., {{\"O}berg}, K.~I., {Wilner}, D.~J., \& {Rosenfeld}, K.~A. 2013,
  \apjl, 765, L14

\bibitem[{{Rab} {et~al.}(2018){Rab}, {G{\"u}del}, {Woitke}, {Kamp}, {Thi},
  {Min}, {Aresu}, \& {Meijerink}}]{rab2018}
{Rab}, C., {G{\"u}del}, M., {Woitke}, P., {et~al.} 2018, \aap, 609, A91

\bibitem[{{Remijan} {et~al.}(2005){Remijan}, {Hollis}, {Lovas}, {Plusquellic},
  \& {Jewell}}]{remijan2005}
{Remijan}, A.~J., {Hollis}, J.~M., {Lovas}, F.~J., {Plusquellic}, D.~F., \&
  {Jewell}, P.~R. 2005, \apj, 632, 333

\bibitem[{{Rimmer} {et~al.}(2012){Rimmer}, {Herbst}, {Morata}, \&
  {Roueff}}]{rimmer2012}
{Rimmer}, P.~B., {Herbst}, E., {Morata}, O., \& {Roueff}, E. 2012, \aap, 537,
  A7

\bibitem[{{Rosenfeld} {et~al.}(2013){Rosenfeld}, {Andrews}, {Wilner},
  {Kastner}, \& {McClure}}]{rosenfeld2013}
{Rosenfeld}, K.~A., {Andrews}, S.~M., {Wilner}, D.~J., {Kastner}, J.~H., \&
  {McClure}, M.~K. 2013, \apj, 775, 136

\bibitem[{{Ruaud} {et~al.}(2016){Ruaud}, {Wakelam}, \& {Hersant}}]{ruaud2016}
{Ruaud}, M., {Wakelam}, V., \& {Hersant}, F. 2016, \mnras, 459, 3756

\bibitem[{{Shakura} \& {Sunyaev}(1973)}]{shakura1973}
{Shakura}, N.~I., \& {Sunyaev}, R.~A. 1973, \aap, 24, 337

\bibitem[{Sutherland(2016)}]{sutherland2016}
Sutherland, J.~D. 2016, Angewandte Chemie International Edition, 55, 104.
\newblock \url{https://onlinelibrary.wiley.com/doi/abs/10.1002/anie.201506585}

\bibitem[{{van der Walt} {et~al.}(2011){van der Walt}, {Colbert}, \&
  {Varoquaux}}]{numpy_article}
{van der Walt}, S., {Colbert}, S.~C., \& {Varoquaux}, G. 2011, Computing in
  Science and Engineering, 13, 22

\bibitem[{{van Dishoeck} \& {Blake}(1998)}]{vandishoeck1998}
{van Dishoeck}, E.~F., \& {Blake}, G.~A. 1998, \araa, 36, 317

\bibitem[{{Vasyunin} {et~al.}(2008){Vasyunin}, {Semenov}, {Henning}, {Wakelam},
  {Herbst}, \& {Sobolev}}]{vasyunin2008}
{Vasyunin}, A.~I., {Semenov}, D., {Henning}, T., {et~al.} 2008, \apj, 672, 629

\bibitem[{{Vasyunin} {et~al.}(2004){Vasyunin}, {Sobolev}, {Wiebe}, \&
  {Semenov}}]{vasyunin2004}
{Vasyunin}, A.~I., {Sobolev}, A.~M., {Wiebe}, D.~S., \& {Semenov}, D.~A. 2004,
  Astronomy Letters, 30, 566

\bibitem[{{Vigren} {et~al.}(2008){Vigren}, {Kami{\'n}ska}, {Hamberg},
  {Zhaunerchyk}, {Thomas}, {Danielsson}, {Semaniak}, {Andersson}, {Larsson}, \&
  {Geppert}}]{vigren2008}
{Vigren}, E., {Kami{\'n}ska}, M., {Hamberg}, M., {et~al.} 2008, Physical
  Chemistry Chemical Physics (Incorporating Faraday Transactions), 10, 4014

\bibitem[{{Wagenblast} \& {Hartquist}(1989)}]{wagenblast1989}
{Wagenblast}, R., \& {Hartquist}, T.~W. 1989, \mnras, 237, 1019

\bibitem[{{Waggoner} \& {Cleeves}(2019)}]{waggoner2019}
{Waggoner}, A.~R., \& {Cleeves}, L.~I. 2019, arXiv e-prints, arXiv:1908.08048

\bibitem[{{Wakelam} {et~al.}(2019){Wakelam}, {Chaietillon}, {Dutrey},
  {Guilloteau}, {Iqbal}, {Coutens}, \& {Majumdar}}]{wakelam2019}
{Wakelam}, V., {Chaietillon}, E., {Dutrey}, A., {et~al.} 2019, \mnras, 484,
  1563

\bibitem[{{Wakelam} {et~al.}(2010){Wakelam}, {Herbst}, {Le Bourlot}, {Hersant},
  {Selsis}, \& {Guilloteau}}]{wakelam2010}
{Wakelam}, V., {Herbst}, E., {Le Bourlot}, J., {et~al.} 2010, \aap, 517, A21

\bibitem[{{Wakelam} {et~al.}(2006){Wakelam}, {Herbst}, \&
  {Selsis}}]{wakelam2006}
{Wakelam}, V., {Herbst}, E., \& {Selsis}, F. 2006, \aap, 451, 551

\bibitem[{{Wakelam} {et~al.}(2016){Wakelam}, {Ruaud}, {Hersant}, {Dutrey},
  {Semenov}, {Majumdar}, \& {Guilloteau}}]{wakelam2016}
{Wakelam}, V., {Ruaud}, M., {Hersant}, F., {et~al.} 2016, \aap, 594, A35

\bibitem[{{Wakelam} {et~al.}(2005){Wakelam}, {Selsis}, {Herbst}, \&
  {Caselli}}]{wakelam2005}
{Wakelam}, V., {Selsis}, F., {Herbst}, E., \& {Caselli}, P. 2005, \aap, 444,
  883

\bibitem[{{Wakelam} {et~al.}(2012){Wakelam}, {Herbst}, {Loison}, {Smith},
  {Chandrasekaran}, {Pavone}, {Adams}, {Bacchus-Montabonel}, {Bergeat},
  {B{\'e}roff}, {Bierbaum}, {Chabot}, {Dalgarno}, {van Dishoeck}, {Faure},
  {Geppert}, {Gerlich}, {Galli}, {H{\'e}brard}, {Hersant}, {Hickson},
  {Honvault}, {Klippenstein}, {Le Picard}, {Nyman}, {Pernot}, {Schlemmer},
  {Selsis}, {Sims}, {Talbi}, {Tennyson}, {Troe}, {Wester}, \&
  {Wiesenfeld}}]{wakelam2012}
{Wakelam}, V., {Herbst}, E., {Loison}, J.~C., {et~al.} 2012, \apjs, 199, 21

\bibitem[{{Whittet}(2010)}]{whittet2010}
{Whittet}, D.~C.~B. 2010, \apj, 710, 1009

\bibitem[{{Williams} \& {Best}(2014)}]{williams2014}
{Williams}, J.~P., \& {Best}, W.~M.~J. 2014, \apj, 788, 59

\end{thebibliography}

\end{document}